\documentclass[aps,prd,groupedaddress]{revtex4}
\usepackage{graphicx}
\usepackage{epsfig}
\usepackage{amsfonts}
\usepackage{amssymb}
\usepackage{epsf}
\usepackage{color}
\newcommand{\insertplot}[5]{\begin{figure}
 \hfill\hbox to 0.05in{\vbox to #5in{\vfill
 \inputplot{#1}{#4}{#5}}\hfill}
 \hfill\vspace{-.1in}
 \caption{#2}\label{#3}
 \end{figure}}
\newcommand{\inputplot}[3]{
 \special{ps: plotfile #1}
}

\newcounter{fig}

\newcommand{\ee}{\end{equation}}
\newcommand{\eea}{\end{eqnarray}}
\newcommand{\be}{\begin{equation}}
\newcommand{\bea}{\begin{eqnarray}}

\usepackage{xcolor}

\begin{document}

\title{Charged and radially excited boson stars (in Anti-de Sitter)}
\author{Yves Brihaye}
\email[]{yves.brihaye@umons.ac.be}
\affiliation{Service de Physique de l'Univers, Champs et Gravitation, Universit\'e de Mons, Mons, Belgium}
\author{Felipe Console}
\email[]{felipe.console@usp.br}
\affiliation{Instituto de F\'isica de S\~ao Carlos, Universidade de S\~ao Paulo, S\~ao Carlos, S\~ao Paulo 13560-970, Brazil}
\author{Betti Hartmann}
\email[]{b.hartmann@ucl.ac.uk}
\affiliation{Department of Mathematics, University College London, Gower Street, London, WC1E 6BT, UK}

\date{\today}

\begin{abstract}
We study charged and radially excited boson stars in both asymptotically flat as well as asymptotically Anti-de Sitter space-time. We demonstrate that two different types of radially excited solutions exist~: one that persists in the linear limit of small scalar fields, in which analytical arguments suggest the existence of these solutions, and one that appears only in the highly non-linear regime of the model. We also demonstrate that the formation of wavy scalar hair discussed previously for black holes and boson stars in asymptotically flat space-time persists for asymptotically Anti-de Sitter space-time. 

\end{abstract}

\maketitle


\section{Introduction}
Anti-de Sitter (AdS) -- together with Minkowski and de Sitter (dS) -- space-time plays an important role amongst known exact solutions of the Einstein equation. All three space-times are maximally symmetric, i.e. admit the largest possible amount of Killing vectors. Minkowski, dS and AdS possess constant vanishing, positive and negative curvature, respectively, with the Ricci scalar being proportional to the cosmological constant. Amongst these three space-times, AdS is special as it possesses a conformal boundary, i.e. a boundary that is in causal contact with the interior of the space-time. As such, physics in the interior can be connected to (albeit different) physics on the boundary. The Anti-de Sitter/Conformal Field Theory (AdS/CFT) correspondence \cite{Maldacena:1997re} (or more generally the gauge-gravity correspondence \cite{gauge_gravity})  as a prediction rooted in String Theory
uses exactly this idea: a gravity theory in $d$-dimensional AdS is dual to a conformal field theory on the $(d-1)$-dimensional boundary of AdS. The duality also predicts that there exists a weak-strong coupling duality
between the theories in the bulk and on the boundary.
Consequently, solutions to (weakly coupled) gravitational theories in $d$-dimensional asymptotically AdS (aAdS) have been used to make predictions about strongly coupled field theories in $(d-1)$ dimensions. One of the main examples
that has been widely studied are models for high-temperature superconductivity \cite{hhh}. Using bulk black holes
allows to study the temperature dependence of the phase transition's order parameter, while phase transitions
can also be modelled using solitonic objects \cite{Horowitz:2010jq}. The latter have no temperature ascribed to them, but phase transitions appear e.g. when varying the chemical potential. 
Most discussions of these so-called {\it holographic superconductors} have been done using a scalar field in the bulk and consequently interpreting the value of this scalar field on the conformal boundary as the expectation value of an operator in the dual field theory with dimension equal to the power of the fall-off of the scalar field in aAdS. In the language of condensed matter, these superconductors would correspond to $s$-wave superconductors. 

Scalar fields are well motivated in Particle Physics models as well as effective descriptions of collective phenomena. 
$Q$-balls are lumps of complex scalar field carrying a conserved Noether charge that results from the invariance
of the model under a U(1) gauge transformation \cite{Coleman}. They possess a scalar field that oscillates 
with constant frequency such that the energy-momentum tensor (and consequently the space-time) stays static.  These objects appear e.g. in supersymmetric extensions of the Standard Model \cite{Kusenko} and need specific scalar field potentials (with attractive and repulsive terms) in order to exist. In \cite{Copeland} an effective exponential potential that appears in gauge-mediated supersymmetric breaking has been discussed in this context. 
Boson stars \cite{Kaup,Friedberg, Jetzer,Schunck} are the curved space-time equivalent of $Q$-balls. Other than their flat space-time counterparts, they do not require a higher order scalar field potential, but exist for
a non-self-interacting and massive scalar field. Boson stars also exist in aAdS space-time \cite{radu}, but now
-- in contrast to the asymptotically flat space-time case -- are no longer exponentially localized, but 
possess a power-law fall-off for the scalar field function. The power of this fall-off is determined by the mass
and the angular frequency of the scalar field. The value of the scalar field on the conformal boundary can then be interpreted as the expectation value of a dual operator in the corresponding boundary field theory, e.g.
as a glueball condensate \cite{Hartmann_Riedel}.

Interesting new phenomena appear when gauging the U(1) symmetry, i.e. when electrically charging the boson star. This has been studied for
a 4th-order scalar field potential \cite{cbs1},  a $V$-shaped potential \cite{cbs2,cbs3}, an exponential potential \cite{cbs4}, a 6th order potential \cite{gauged_BS_power_law} as well as for a massive scalar field with no self-interaction \cite{cbs5}, respectively. When the gauge interaction becomes strong, charged boson stars cease to exist, which is related to the fact that the individual constituents, the scalar bosons, that make up the star are now also electrically charged and hence the electric repulsive force becomes important. 

In this paper, we study charged boson stars in asymptotically flat and AdS space-time, respectively. In the former case, which
has been discussed in the literature previously, we show that radially excited charged boson stars exist
and that these have very different qualitative properties as compared to the unexcited boson stars.
In particular, we demonstrate that two disconnected branches of solutions exist.

\section{Set-up}

We consider a U(1)-charged complex scalar field, $\Psi$, minimally coupled to gravity with negative cosmological constant $\Lambda$ and a scalar field potential $\mathcal{U}$. The action is given by
\begin{equation}
S =  \int d^4 x \sqrt{-g} \left(\frac{\mathcal{R}}{16 \pi G} - \frac{\Lambda}{8 \pi G} -\left( D_{\mu}\Psi\right) (D^{\mu}\Psi)^{*} - \frac{1}{4} F_{\mu\nu}F^{\mu\nu} - \mathcal{U}\left( |\Psi|^2\right) \ \right)
\end{equation}
where $\mathcal{R}$ is the Ricci scalar, $G$ is the Newton gravitational constant, $D_{\mu} = \nabla_{\mu} + i q A_{\mu}$ is the gauge covariant derivative, $A_{\mu}$ the U(1) gauge field, $F_{\mu\nu} = \partial_{\mu}A_{\nu} - \partial_{\nu}A_{\mu}$ the field strenght tensor and $q$ the scalar charge.

The equations of motion are obtained by varying the action with respect to metric and the matter fields and read, respectively~:
\begin{equation}\label{syseqads}
        G_{\mu\nu} + \Lambda g_{\mu\nu} = 8 \pi G T_{\mu\nu}\hspace{.2cm}, \hspace{1cm}    \nabla_{\mu}F^{\nu\mu} = i q \left( \Psi^{*} \left(D^{\nu}\Psi \right) - \Psi \left( D^{\nu} \Psi \right)^{*} \right) \equiv q j^{\nu}  \hspace{.2cm}, \hspace{1cm}         D_{\mu} D^{\mu} \Psi = \frac{d \mathcal{U}}{d |\Psi|^2} \Psi \hspace{.2cm}.
\end{equation}
$G_{\mu\nu}$ is the Einstein tensor and the energy-momentum tensor, $T_{\mu\nu}$, can be written as the sum of the energy momentum-tensor for the scalar field, $T_{\mu\nu}^{(\Psi)}$, and for the electromagnetic field, $T_{\mu\nu}^{(EM)}$ as follows~:
\begin{equation}
T_{\mu\nu} = T_{\mu\nu}^{(\Psi)} + T_{\mu\nu}^{(EM)}
\end{equation}    
with
\begin{equation}
T_{\mu\nu}^{(\Psi)}  =   \left( D_{\mu}\Psi \right)\left( D_{\nu}\Psi \right)^{*} + \left( D_{\mu}\Psi \right)^{*}\left( D_{\nu}\Psi  \right) - g_{\mu \nu} \left( \left( D_{\alpha}\Psi \right)\left( D^{\alpha}\Psi \right)^{*} + \mathcal{U}  \right) \hspace{.2cm},  \hspace{1cm}   T_{\mu\nu}^{(EM)}  =  F_{\mu \alpha} F_{\nu}^{ \hspace{.2cm} \alpha} -\frac{1}{4} g_{\mu\nu} F_{\alpha \beta} F^{\alpha\beta} \hspace{.2cm}.
\end{equation}
We consider a scalar self-interacting potential of the form~:
\begin{equation}
\mathcal{U}\left( |\Psi|^2 \right) = \mu^2 \eta^2 \left( 1 - \exp \left( -\frac{|\Psi|^2}{\eta^2} \right)  \right) \hspace{.2cm},
\end{equation}
where $\mu$ is the mass of the boson and $\eta$ is an energy scale. This potential has been considered previously in the construction of Q-balls and boson stars in AdS \cite{Copeland, Hartmann_Riedel}. 

In the following the space-time is chosen to be static and spherically symmetric with adapted coordinates $\{t, r, \theta, \varphi\}$ such that $\xi = \partial_t$ and $\eta = \partial_{\varphi}$ are a timelike and spacelike Killing vector, respectively. On the other hand, the Ansatz for the matter fields is non-static since, although static solutions are not forbidden by Derrick's theorem when gauge fields are involved, static gauge fields with vanishing $A_t$ component lead to magnetic fields only and these cannot be spherically symmetric in Abelian gauge field theory (there are no monopole terms). However, we will choose an Ansatz that will lead to a static energy-momentum tensor to be consistent with a static space-time. The Ansatz for the metric, scalar field $\Psi$ and U(1) gauge field $A_{\mu}$ then reads~: 
\begin{equation}
    {\rm d}s^2= - N(r) \sigma(r)^2 {\rm d}t^2 + \frac{1}{N(r)} {\rm d}r^2 + r^2 {\rm d}\theta^2 + r^2 \sin^2 \theta {\rm d}\varphi^2 \hspace{.1cm}, \hspace{.4cm} \Psi = \Psi(t, r) = e^{-i \omega t} \psi\left( r \right) \hspace{.1cm}, \hspace{.4cm} A_{\mu} {\rm d}x^{\mu}= V(r) {\rm d}t \hspace{.2cm},
 \end{equation}
where we will use $N(r) = 1 -2m(r)/r$ with $m(r)$ a mass function.
Replacing the above Ansatz into (\ref{syseqads}) we obtain a system of four coupled non-linear ordinary differential equations, which read~:
\begin{equation}\label{charged_AdS_eq1}
    m' = \frac{1}{2}r^2\Lambda + 4 \pi G r^2 \left( \mathcal{U} + \frac{\left( \omega - q V \right)^2}{N \sigma^2} \psi^2 + \frac{(V')^2}{2 \sigma^2}  + N (\psi')^2\right) \hspace{1cm} \sigma' = \frac{8 \pi G r}{N^2 \sigma } \left( \left( \omega - q V \right)^2 \psi^2 + N^2 \sigma^2 \left( \psi' \right)^2  \right)
\end{equation}
\begin{equation}\label{charged_AdS_eq2}
    \psi'' =  -\left( \frac{N'}{N} + \frac{\sigma'}{\sigma}  + \frac{2}{r} \right) \psi' - \frac{\left( \omega - q V \right)^2}{N^2 \sigma^2} \psi + \frac{1}{N} \frac{d \mathcal{U}}{d |\Psi|^2} \psi \hspace{1cm} V'' = - \left( \frac{2}{r} -\frac{\sigma'}{\sigma} \right) V' - \frac{2q\left( \omega - q V \right)}{N} \psi^2 \ .
\end{equation}
The above equations obviously depend only on the gauge invariant combination $\omega-qV$, which also appears in the expression for the globally conserved Noether charge that reads~:
\begin{equation}
\label{eq:noether}
Q_{N} =  \int_{0}^{\infty} {\rm d}r \hspace{.05cm}\frac{2 r^2 \psi^2}{N \sigma } \left( q V - \omega \right) \hspace{.2cm}.
\end{equation}
$Q_N$ has often been interpreted as the number of bosonic particles of mass $\mu$ that make up the boson star.
While in the ungauged case, next to the gravitational attraction, only the scalar field potential is responsible for  repulsion between these individual constituents, there is an additional repulsive electric force acting in the gauged case. Hence, the properties of electrically charged boson stars are similar to those of the ungauged stars as long as the electric field is small, but deviate significantly when the gauge interaction becomes important.  

In order to find globally regular solutions in an asymptotically AdS space-time (aAdS), we need to impose appropriate boundary conditions. Close to the origin $r=0$ the equations of motion (\ref{charged_AdS_eq1}) and (\ref{charged_AdS_eq2}) imply that
\begin{equation}\label{bc_charged_bs_1}
m( 0 ) = 0 \hspace{.1cm}, \hspace{1cm}  \hspace{1cm} \psi'(0) = 0 \hspace{.1cm},  \hspace{1cm} V'(0) = 0 \hspace{.1cm}, 
\end{equation}
while $\psi( 0 ) \equiv \psi_0$, $\sigma(0) \equiv \sigma_{0}$
and $V(0)\equiv V_0$ are {\it a priori} free parameters. However, we can use the gauge freedom to fix $V(0)=0$. The values $\psi_0$
and $\sigma_{0}$ will, on the other hand, be determined numerically. 

Asymptotically, the leading terms of the metric functions are~:
\begin{equation}
N(r \rightarrow \infty	) \sim 1 - \frac{\Lambda}{3} r^2  \hspace{.1cm},  \hspace{1cm} \sigma(r \rightarrow \infty ) \sim 1  \hspace{.2cm}.
\end{equation}
from which we can derive the behaviour of the matter field functions. This becomes~:
\begin{equation}
V(r \rightarrow \infty ) \sim \Phi - \frac{Q_{e}}{r}\ \ \ , \ \ 
\psi(r \rightarrow \infty ) \sim   \frac{\psi_{+}}{r^{\Delta_{+}}} + \frac{\psi_{-}}{r^{\Delta_{-}}}  \hspace{.2cm}, \hspace{1cm} \Delta_{\pm} = \frac{3}{2} \pm  \frac{1}{2}\sqrt{9 -\frac{12 \mu^2}{\Lambda}} \ \ , 
\end{equation}
where $\Phi$ is a constant that we will often refer to as ``the chemical potential'' and 
$Q_e$ is the electric charge of the solution that can be shown to be related to the Noether charge by $Q_e=eQ_N$. In fact, with our choice $V(0)=0$, $\Phi$ denotes the electric potential difference between the origin and infinity. 

Note that, different from the asymptotically flat space-time case, where $\psi$ falls of exponentially
\begin{equation}
    \psi(r \rightarrow \infty ) \sim \frac{\exp\left(-\sqrt{\mu^2 - (\omega-q\Phi)^2}r\right)}{r} \ \ \ \ \ \ \ \ \  \ \ (\Lambda=0) \ ,
\end{equation}
the scalar field now has a power-law fall-off as long as $\mu^2 > 3 \Lambda/4$. 
In the following, we will choose $\mu^2 > 0 $, and since $\Lambda < 0$, this implies that $\Delta_{-}$ is negative and the corresponding solution to $\psi$ would diverge as $r \rightarrow \infty$. For this reason we set $\psi_{-} = 0$ and consider only the regular solution at infinity, $\psi \sim \psi_+/r^{\Delta_+}$. $\psi_{+}$ is a constant that can then be interpreted as the expectation value of a ``condensate'' in the dual description on the conformal boundary.  

Finally, the mass $M$ of the solution can be read of from the asymptotic behaviour of the mass function $m(r)$, which behaves as \cite{radu}~:
\begin{equation}
    m(r \rightarrow \infty ) \sim M + \frac{\Lambda r^3}{6} -\frac{2 \pi G Q^2}{r}+ n_1 r^{-2\Delta_++3} \ ,
\end{equation}
where $n_1$ is a constant depending on $\Lambda$ and we need to subtract the diverging part for $\Lambda \neq 0$, which - more formally - would be done using the counterterm approach.

\subsection{Radially excited solutions and gauged oscillons}
\label{subsection:oscillons}
Assuming $\psi$ to be small we find the linearized scalar field equation
\begin{equation}
\label{eq:linear}
\frac{{\rm d}}{{\rm d}z}\left(z^2 f \frac{{\rm d}\psi}{{\rm d}z}\right) = \left(\mu^2 - \frac{\Omega^2}{f}\right) \ell^2 z^2 \psi \ \ , \ \ f=1+z^2 \ ,     
\end{equation}
where we have defined $z=r/\ell$ with $\ell=\sqrt{-3/\Lambda}$ the AdS radius and $\Omega = \omega - q\Phi$. The solutions to this equation are so-called {\it oscillons} and have been discussed in the ungauged case before (see e.g. \cite{Bizon:2011gg, Dias:2012tq,Brihaye:2014bqa}). 
Since the scalar field sources the U(1) gauge field only via a current proportional to $\psi^2$, the gauge field equation can be integrated to $V(r)\sim \Phi - Q_e/r$ in the limit of $\psi \ll 1$. The leading term of the gauge field inserted into the scalar field equation is then $V\sim \Phi$, i.e. $V$ equal to its chemical potential. 
Now, we will introduce
\begin{equation}
      y = \frac{1}{1+z^2}   \ \ , \ \   \psi = y^p  F(y) \ \ ,
\end{equation}
to rewrite (\ref{eq:linear}). This gives a hypergeometric equation of the form
\begin{equation}
\label{eq:hypergeometric}
         y(y-1) \frac{{\rm d}^2 F}{{\rm d}y^2} 
				+ ((a + b +1)y - c ) \frac{{\rm d} F}{{\rm d}y} +  ab  F = 0
				\end{equation}
with  
\begin{equation}
a = p - \frac{\ell \Omega}{2} \ \ , \ \ b = p + \frac{\ell \Omega}{2} \ \ , \ \ c = 2p+1 - \frac{3}{2}
\end{equation}
and the parameter $p$ is determined by
\begin{equation}
      4 p^2 - 6p - \ell^2 \mu^2 = 0 \ \ \longrightarrow \ \   p_{\pm} = (3 \pm \sqrt{9 + 4 \ell^2 \mu^2})/4 \ .
\end{equation}
The general solution to (\ref{eq:hypergeometric}) is given in terms of the hypergeometric function $\phantom{X}_2 F_1(a,b,c;y)$ and reads
\begin{equation}
	               F(y) = C_1 \phantom{X}_2 F_1(a,b,c; y) + C_2 y^{1-c} \phantom{X}_2 F_1(a-c+1,b-c+1 2-c; y) \ ,
\end{equation}
where $C_i$, $i=1,2$ are integration constants. In general, this solution is not regular for $y \to 0$ unless
$a=-k$, $k\in \mathbb{N}$. This leads to the following requirement for $\Omega$:
\begin{equation}
    \Omega = \frac{2(p+k)}{\ell} \ \ \ , \ \ \ k\in \mathbb{N} \ ,
\end{equation}
i.e. $\Omega$ is quantised. Moreover, the solutions indexed by $k$
possess $k$ zeros. These provide the {\it gauged oscillon} basis
that describes the solutions to the linearized equation (\ref{eq:linear}). This oscillon basis has been
discussed previously in the case of massless scalar fields \cite{Bizon:2011gg,Dias:2012tq} as well
as including the mass of the scalar field \cite{Brihaye:2014bqa}. 

We hence expect that solutions with
zeros in the scalar field function exist also in the non-linear case and would interpret them as radially excited solutions. In the following we will demonstrate that these solutions can, indeed, be constructed numerically
and agree with the analysis done above in the linear limit. 
Note that in the asymptotically flat case, $\Lambda=0$, there is no equivalent
linear argument. However, as we will demonstrate below, excited solutions do exist.

\section{Results}
In order to solve the equations numerically it is convenient to rescale as follows
\begin{equation}\label{substitution_bs_charged_ads}
m \rightarrow  \frac{m} {\mu} \hspace{.1cm}, \hspace{.8cm} \Omega \rightarrow \mu \Omega \hspace{.1cm},  \hspace{.8cm}  \Lambda \rightarrow \Lambda \mu^2 \hspace{.1cm}, \hspace{.8cm} \psi \rightarrow \eta \psi  \hspace{.1cm} , \hspace{.8cm} V \rightarrow \eta V \hspace{.1cm},
\end{equation}
and define the following dimensionless quantities~:
\begin{equation}\label{substitution_bs_charged_ads2}
x =  \frac{r}{\mu} \hspace{.1cm}, \hspace{.8cm}  e = \frac{\eta q}{\mu} \hspace{.1cm}, \hspace{0.8cm} \alpha = 4 \pi G \eta^2 \ .
\end{equation}

Solitonic solutions to the model given here have been previously constructed in the asymptotically flat case \cite{cbs4} (see also  \cite{gauged_BS_power_law} for a very similar study, albeit with a 6th order scalar potential) as well as in the ungauged case in aAdS \cite{Hartmann_Riedel}.
Here, we will be interested in the influence of the charge on asymptotically Anti-de Sitter (aAdS) boson stars
as well as radially excited solutions in both aAdS ($\Lambda < 0$) and asymptotically flat ($\Lambda=0$) space-time.

\subsection{Unexcited solutions}

As discussed above, we expect the model to possess solutions with nodes in the scalar field function.
Here, we first discuss the solutions with no nodes, i.e. the solutions that are not radially excited and will frequently refer to them as $k=0$ solutions. 

\subsubsection{$\Lambda=0$}

\begin{figure}[h!]
\begin{center}
{\label{cc1}\includegraphics[width=8cm]{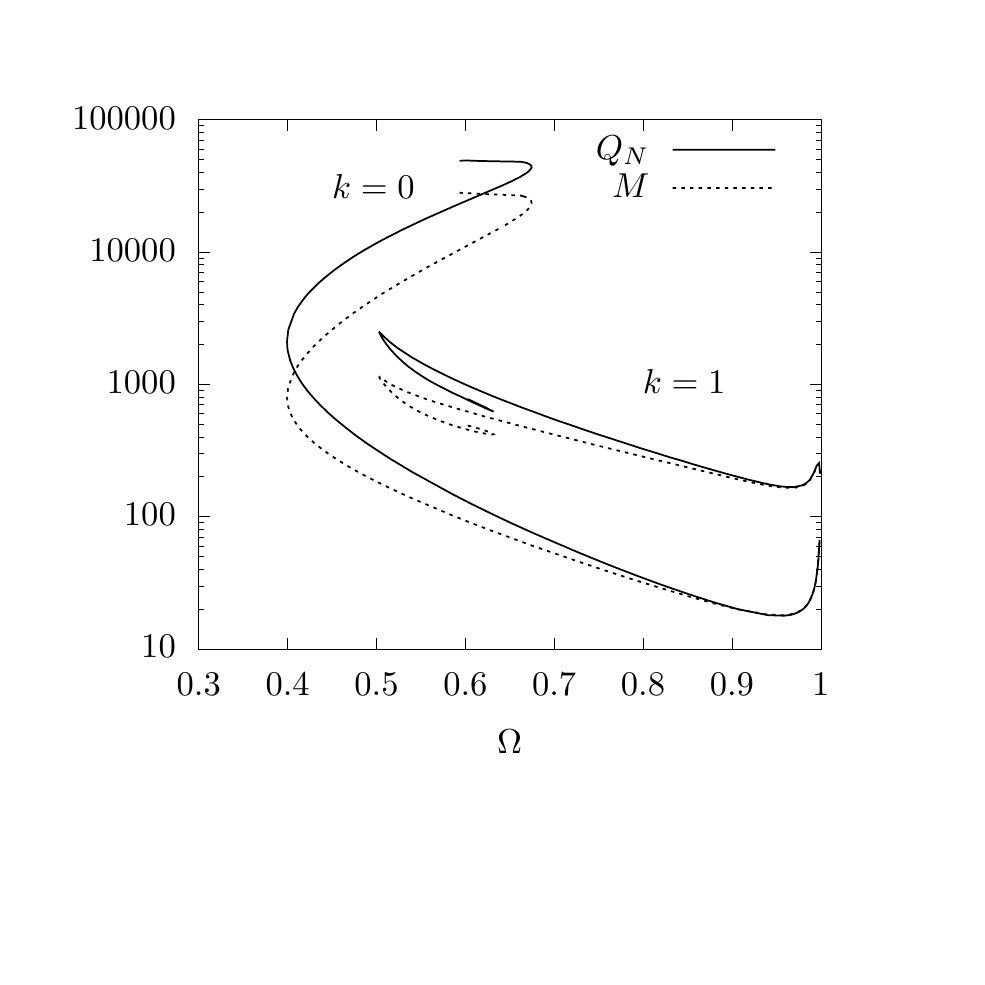}}
{\label{ss0}\includegraphics[width=8cm]{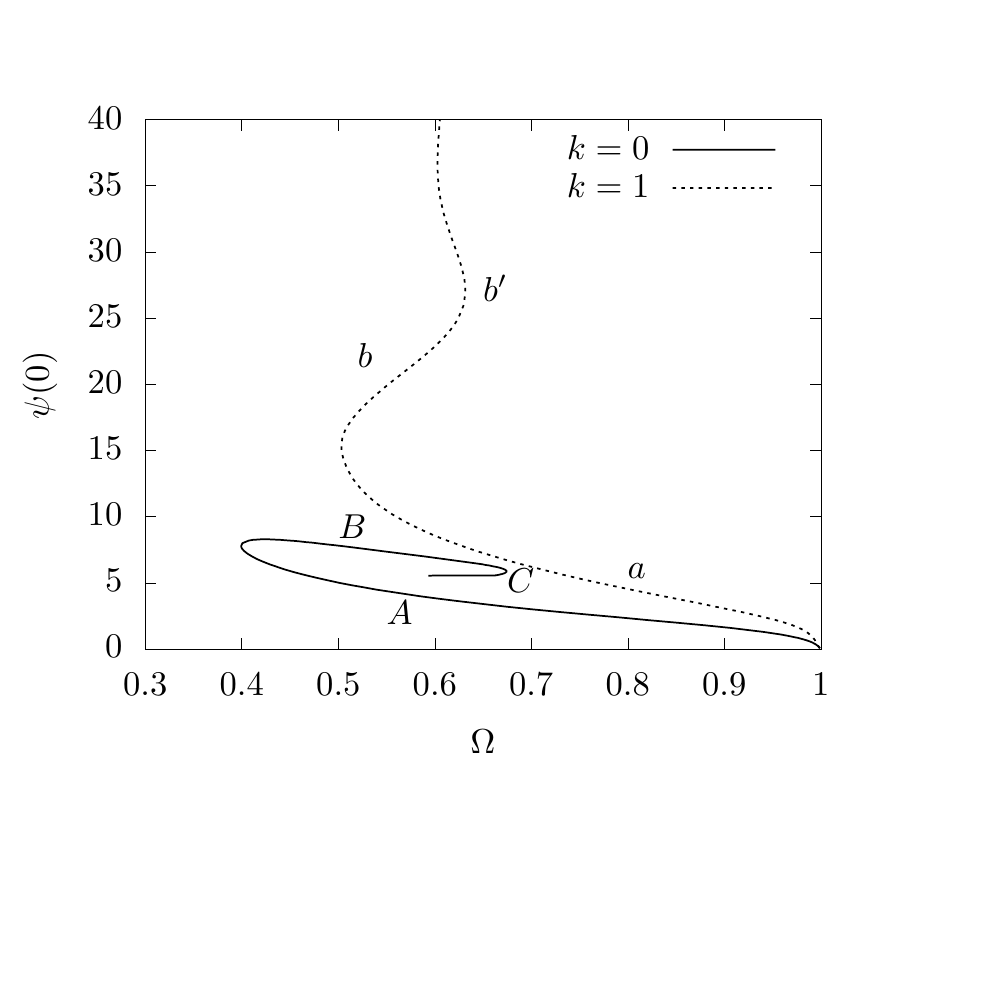}}
\vspace{-2cm}
\caption{{\it Left}: The dependence of the mass $M$ (dashed) and Noether charge $Q_N$ (solid) on $\Omega$ for unexcited boson stars ($k=0$) as well as boson stars with one node in the scalar field function ($k=1$), respectively. Here we have chosen $\alpha =0.0012$, $e=0.02$, $\Lambda=0$.
{\it Right}: The value of the scalar field at the origin, $\psi(0)$, in function of $\Omega$ for the same solutions. $A$, $B$, $C$ indicate the branches for $k=0$ (solid), while $a$, $b$, $b^{\prime}$ refer to those for the $k=1$ solutions (dashed).
\label{fig:flat_unexcited}
}
\end{center}
\end{figure}

Very similar to what has been discussed in \cite{gauged_BS_power_law} for a power-law scalar field potential, we find that also for the exponential scalar field potential three branches of solutions exist when plotting the diverse quantities versus  $\Omega = \omega - e \Phi$ and choosing $\alpha$ and $e$ appropriately. In the following, we will refer to these branches
as branches $A$, $B$ and $C$, respectively, where branch $A$ corresponds to the branch of lowest mass solutions for a given $\Omega$ and exists in the limit
$\Omega \to 1$. In this limit the scalar field $\psi(r)\equiv 0$, while the mass $M$ and Noether charge $Q_N$ both approach a non-vanishing value. 
In Fig. \ref{fig:flat_unexcited} we show $M$ and $Q_N$ (left) as well as the value of the scalar field function at the origin, $\psi(r)$, (right) in function of $\Omega$ for $\alpha=0.0012$ and $e=0.02$. The three branches for the unexcited solution are clearly visible and we note that on all branches we have $M < Q_N$ indicating that the solutions on all branches are stable with respect to the decay into $Q_N$ individual bosons of mass $\mu\equiv 1 $. Moreover, we find that $M\rightarrow Q_N$ in the limit $\Omega \rightarrow 1$. The dependence of $\psi(0)$ on $\Omega$
(see Fig. \ref{fig:flat_unexcited} (right)) demonstrates that the central value of the scalar field cannot exceed a finite value. While on branch $A$ the increase of $\psi(0)$ is connected to an increase of the Noether charge $Q_N$, i.e. the number of bosonic particles making up the boson star, this is no longer true for
branch $B$. On this branch, the number of bosonic particles increases, but the central density decreases.
This suggests that more and more bosonic particles are pushed out from the center due to the increased
electromagnetic repulsion that the bosonic particles feel. On branch $C$ finally, the value of $\psi(0)$ as well as the Noether charge $Q_N$ stay more or less constant.

\begin{figure}[h!]
\begin{center}
{\includegraphics[width=8cm]{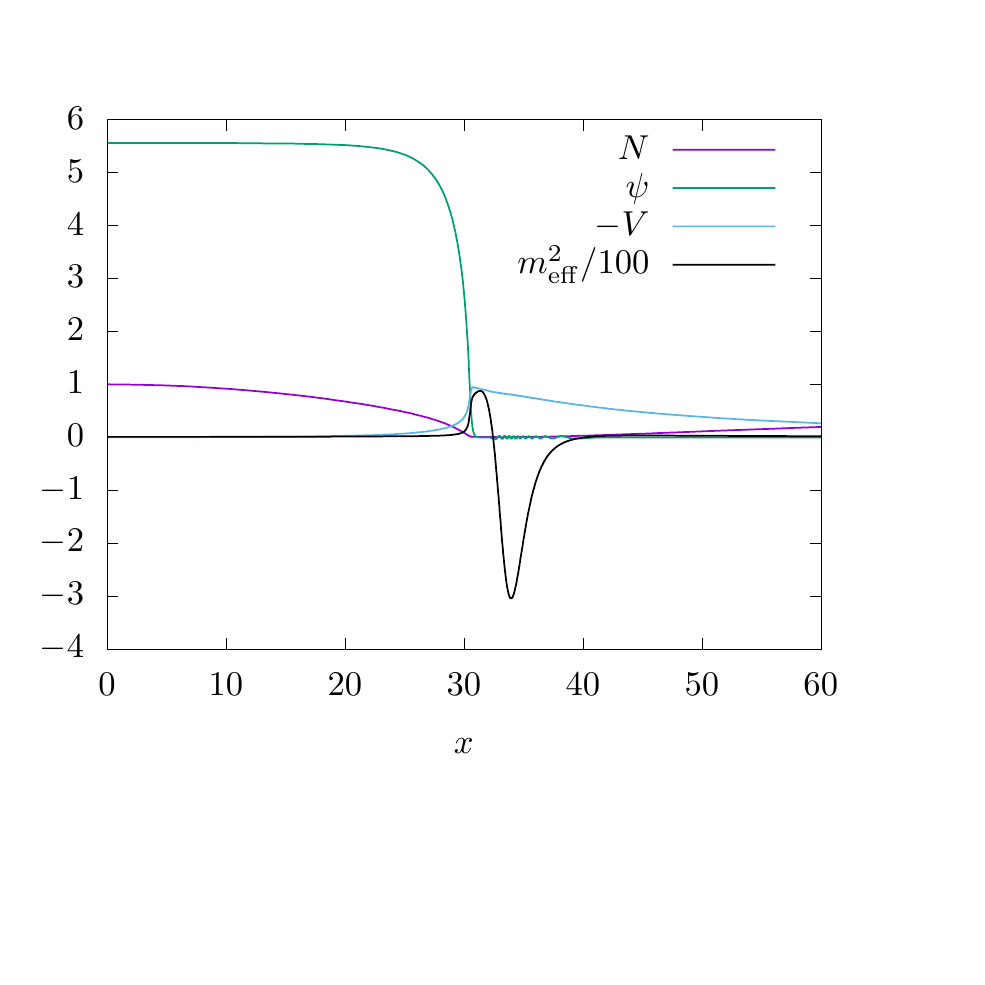}}
{\includegraphics[width=8cm]{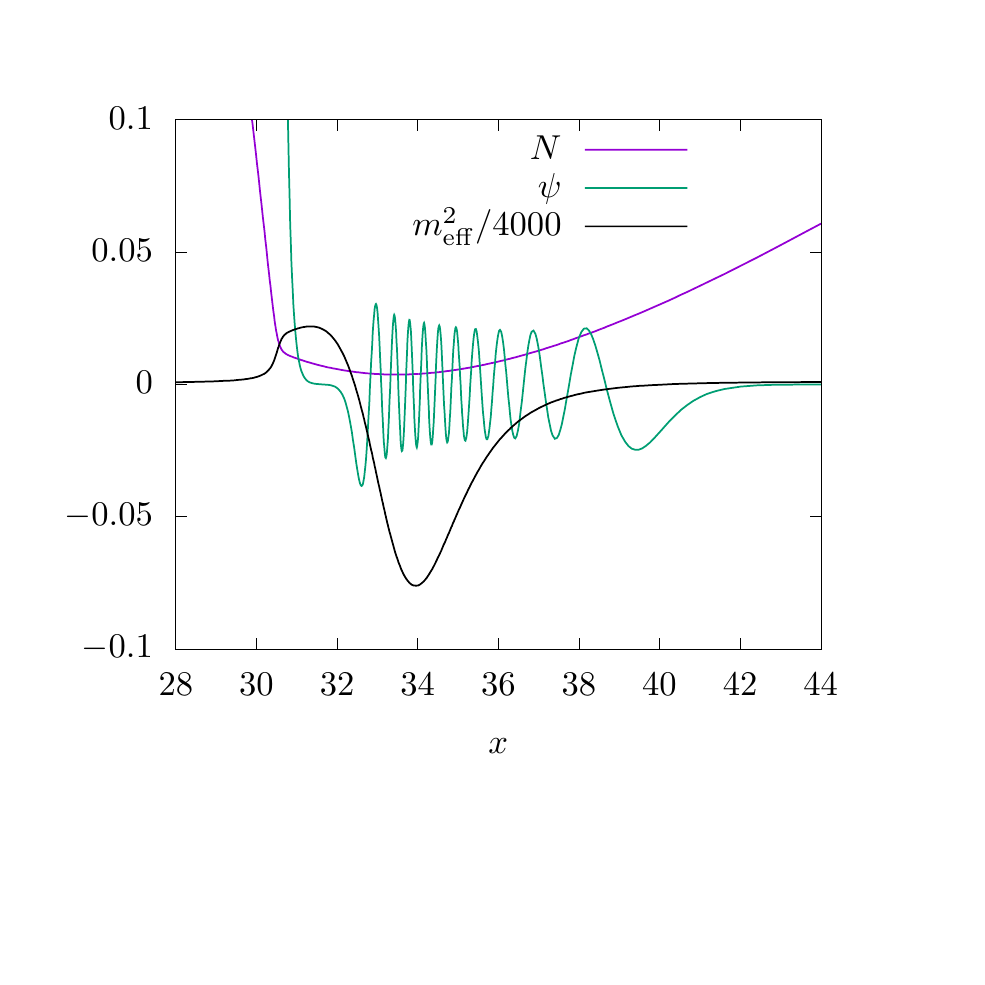}}\\
\vspace{-2cm}

{\includegraphics[width=8cm]{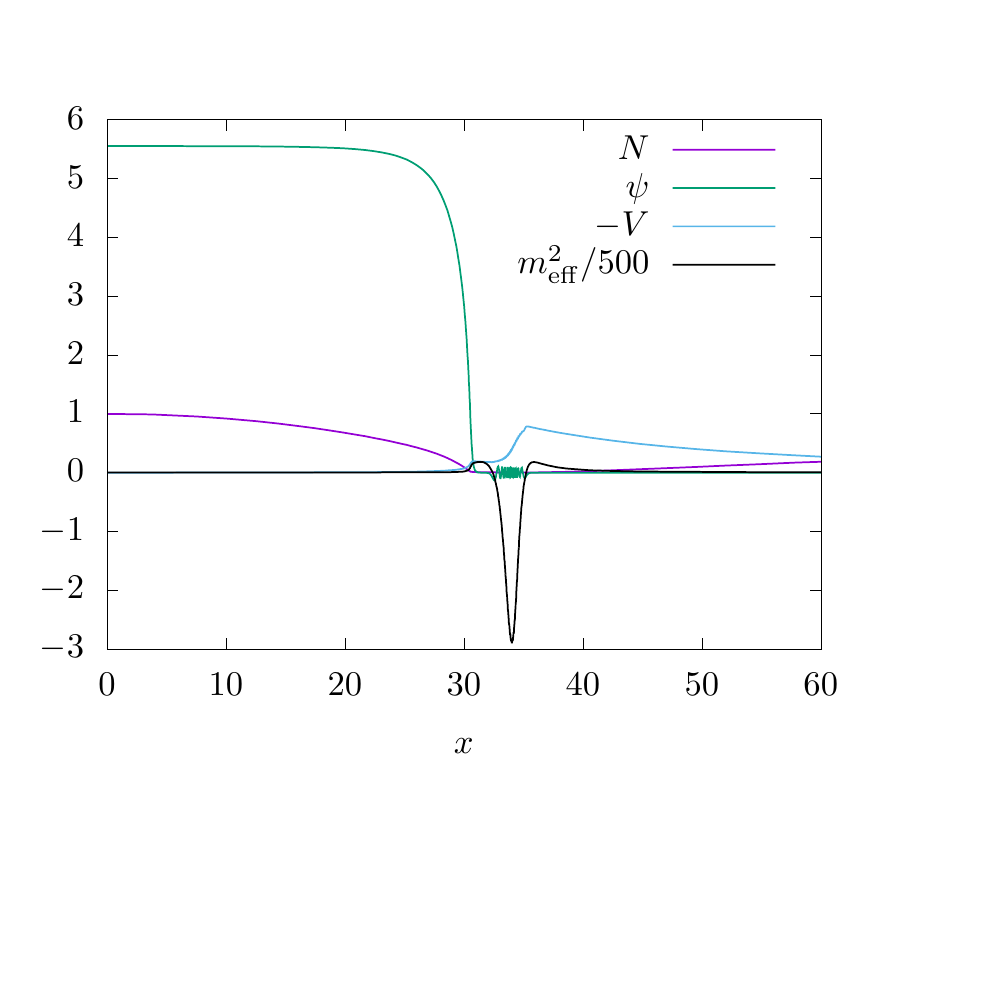}}
{\includegraphics[width=8cm]{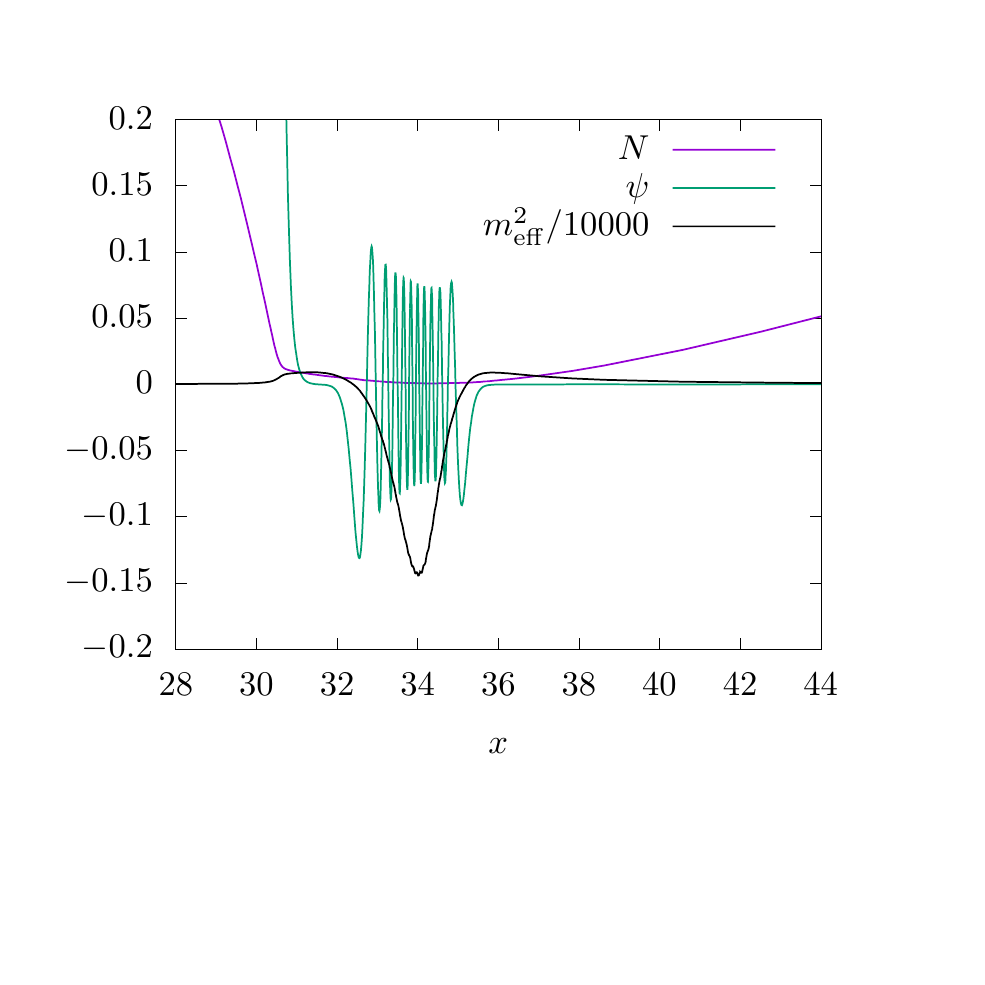}}
\vspace{-2cm}
\caption{We show the profiles of the metric function $N(x)$, of the scalar field function $\psi(x)$, of the electric potential $-V(x)$ (left only)
as well as of the effective mass $m_{\rm eff}^2$ (see (\ref{eq:effectivemass})) for $\alpha = 0.0012$, $e=0.02$ and $\omega = 0.001$ (upper row) as well as $\omega = 0.0002$ (lower row). 
The right figures in each row show a zoom into the $x$-interval where $\psi(x)$ oscillates. 
\label{fig:profile1} 
}
\end{center}
\end{figure}

In order to understand this latter feature, let us discuss in more detail what happens on branch $C$.
Again, this has been discussed previously for a power-law potential in \cite{gauged_BS_power_law}. Here we show that this feature persists for the exponential potential and point out that the interpretation of results
becomes easier when using $\omega$ instead of $\Omega$. As argued in \cite{gauged_BS_power_law}, the scalar
field starts oscillating in regimes where the square of the effective mass $m_{\rm eff}$ defined by (note that $\mu\equiv 1$ due to our rescalings (\ref{substitution_bs_charged_ads}) and (\ref{substitution_bs_charged_ads2}))~:
\begin{equation} 
\label{eq:effectivemass}
m_{\rm eff}^2(x) = \frac{1}{N(x)} - \frac{\left(\omega - e V(x)\right)^2 }{N^2(x) \sigma^2(x)} 
\end{equation}
becomes negative. This is shown in Fig. \ref{fig:profile1} for $\alpha=0.0012$, $e=0.02$ and two different values of $\omega$.
We observe that when decreasing $\omega$ the value of the minimum of the metric function $N(x)$ decreases
and in the limit $\omega\rightarrow 0$ tends to zero. We find e.g. when decreasing $\omega$ from 
$\omega= 0.001$ to $\omega = 0.0002$ that the value of the minimum of $N(x)$ decreases from $\approx 0.004$ to $\approx 0.0007$. At the same time, the number of oscillations in the scalar field function stays constant (equal to ten in our case), but the oscillations increase in amplitude and -- at the same time -- are squeezed to a smaller interval in $x$. This is clearly related to the fact that the effective mass $m^2_{\rm eff}$ is more negative.

\subsubsection{$\Lambda<0$}

\begin{figure}
    \centering
    \includegraphics[scale=0.8]{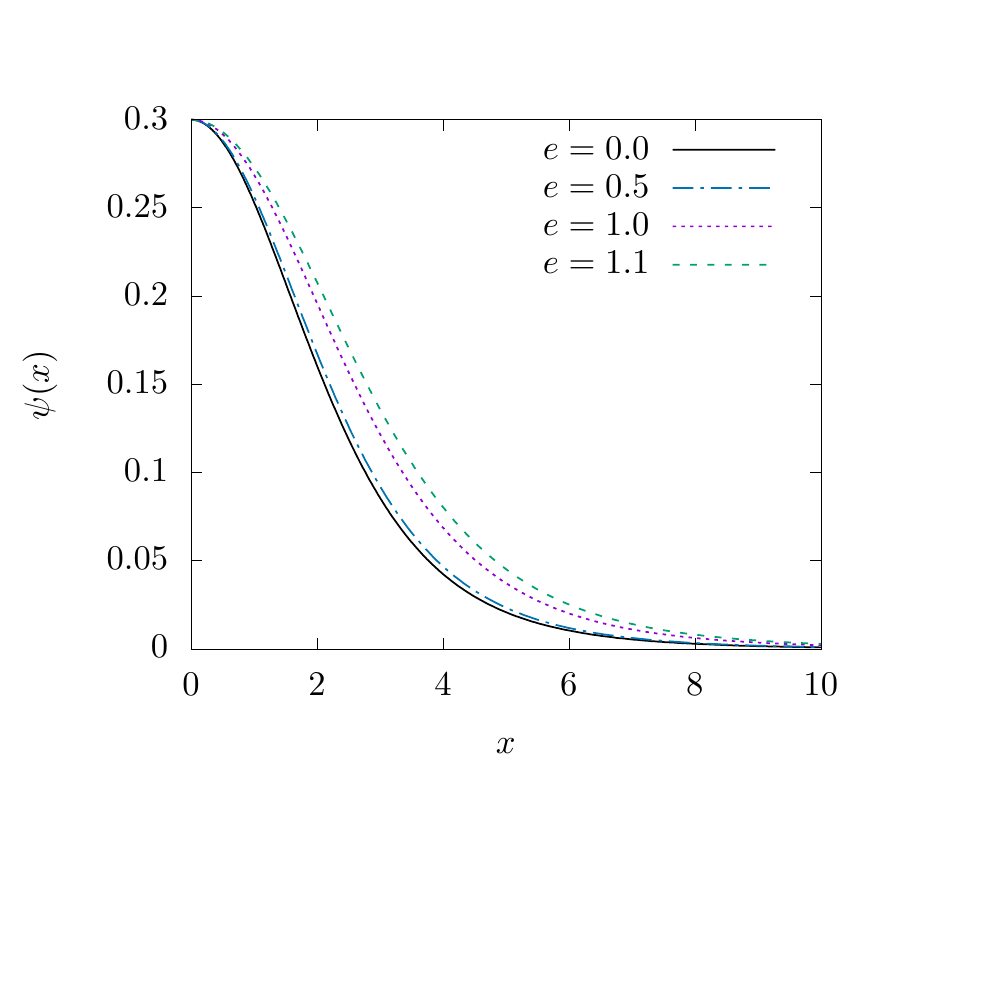}
    \includegraphics[scale=0.8]{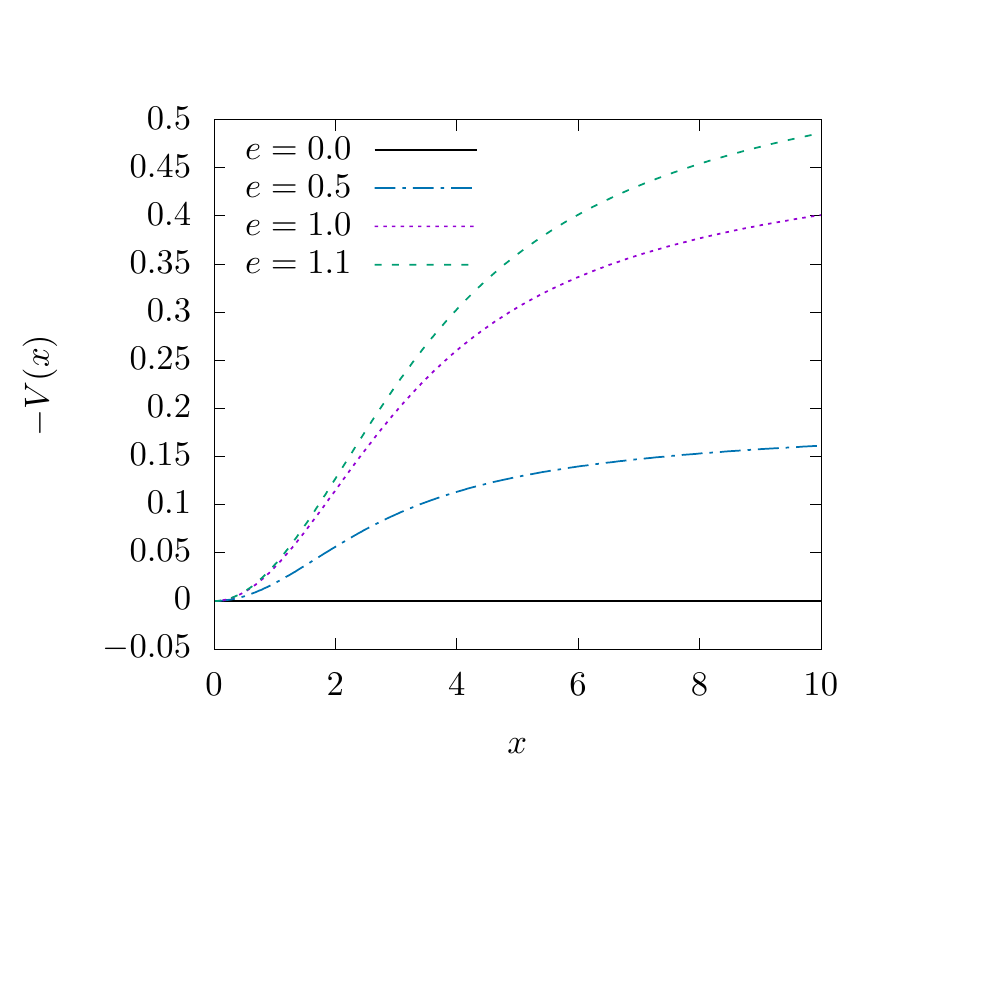}
    \vspace{-2cm}
    \caption{\emph{Left}: We show the scalar field function $\psi(x)$ of the unexcited solution ($k=0)$  for $\alpha = 0$, $\Lambda = -1/6$ and different values of the gauge coupling $e$. \emph{Right}: The corresponding gauge field function $-V(x)$. }
    \label{fig:e_vary_alpha0}
\end{figure}

\begin{figure}
    \centering
    \includegraphics[scale=0.8]{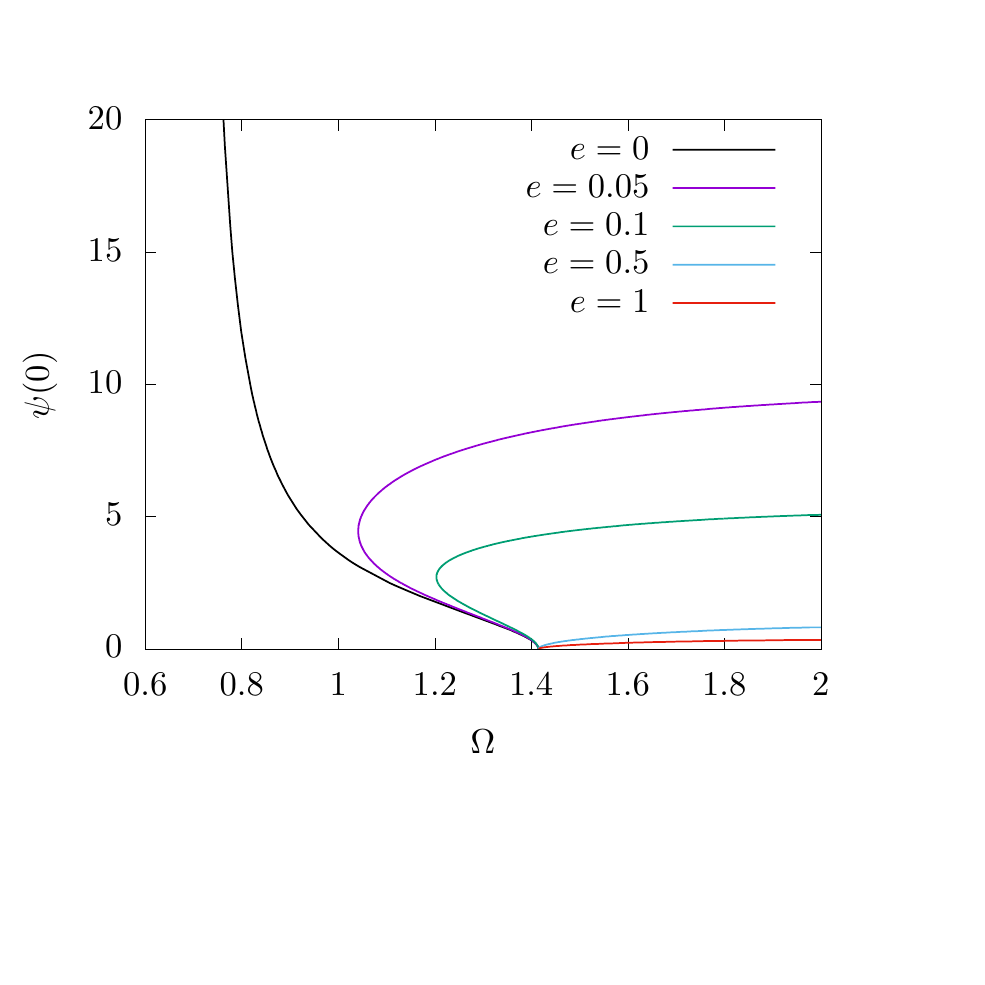}
    \includegraphics[scale=0.8]{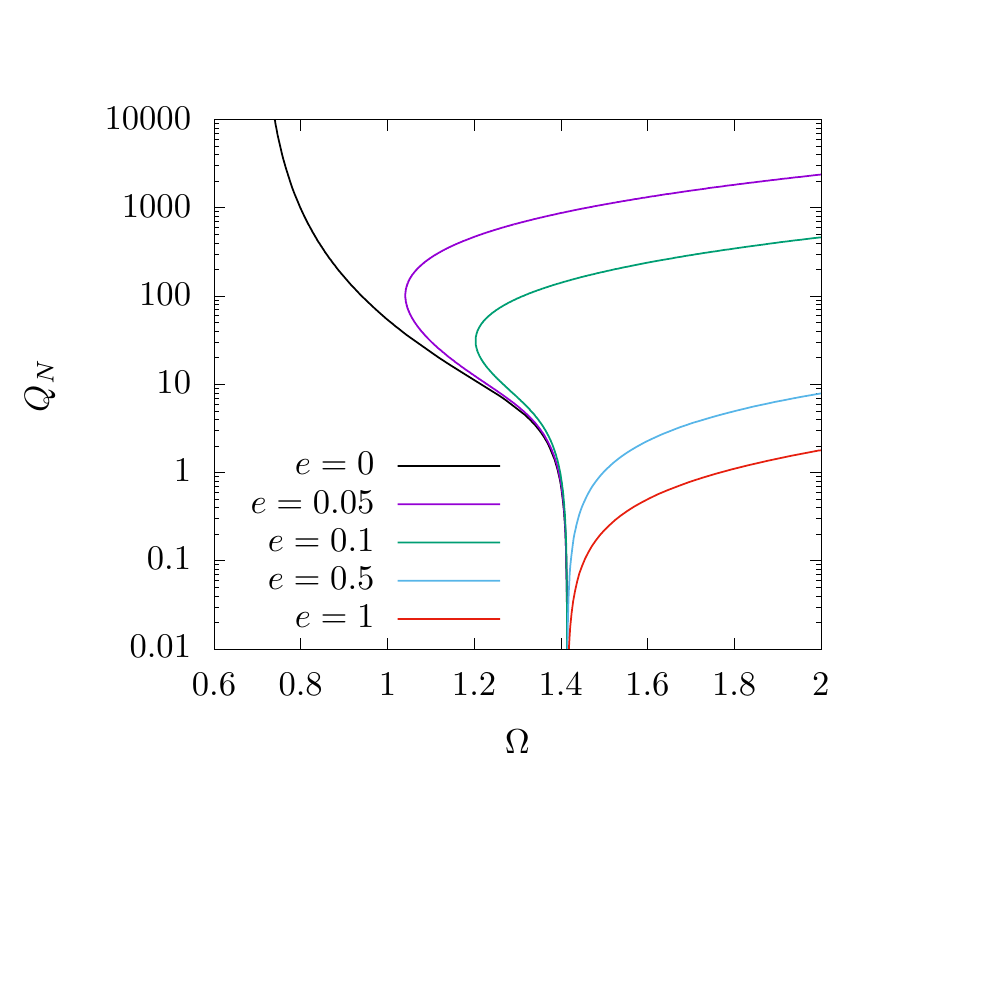}
    \vspace{-2cm}
    \caption{\emph{Left}: We show the value of the scalar field at the origin, $\psi(0)$, as a function of $\Omega$ for $\alpha = 0$, $\Lambda = -1/6$, $k=0$ and different values of $e$. \emph{Right}: The Noether charge $Q_N$ as function of $\Omega$ for the same solutions.}
    \label{fig:psi0_e_alp0} 
\end{figure}

In order to understand the influence of the charge on the solutions, we have first set $\alpha=0$ (such that $N=1-\Lambda x^2/3$, $\sigma\equiv 1$) and studied the behaviour of the solutions when varying $e$.  This is shown in Fig. \ref{fig:e_vary_alpha0} for $e=0, \, 0.5, \, 1, \, 1.1,$ corresponding to 
$\Omega \approx 1.40, \, 1.47, \, 1.76, \, 1.89$, respectively. 
We observe that the scalar field function falls off slower when increasing $e$ which means a larger radius for the solitonic object. This is not surprising noticing that the configuration can be interpreted as being made of $Q_N$ 
scalar and charged bosons, each charged with $e$. Increasing $e$ hence increases the repulsion between the individual constituents and since the space-time does not backreact, there is no attractive force to compensate for the repulsion. Hence, we cannot make a gauged scalar field object possess arbitrarily large scalar field
values (as is, e.g. the case in flat space-time in the ungauged case). This is demonstrated in Fig. \ref{fig:psi0_e_alp0} (left), where we give the value of the scalar field at the origin, $\psi(0)$ in dependence of $\Omega$ for $\alpha=0$, $\Lambda=-1/6$ and different values of $e$. For $e=0$, the value of $\psi(0)$
can become arbitrarily large with the Noether charge $Q_N\rightarrow \infty$ (see Fig. \ref{fig:psi0_e_alp0} (right)). Increasing $e$ from zero, the behaviour is qualitatively different. Solutions exist only down to a minimal
value of $\Omega$, where a second branch of solutions exists that extends backwards in $\Omega$ with increasing
values of $\psi(0)$ and $Q_N$. This means that the frequency $\Omega$ needs to increase in order for the gauged boson star to be composed of more bosonic particles. The larger $e$, the smaller is $\psi(0)$, which again can be explained by the increased electromagnetic repulsion.

\begin{figure}
    \centering
    \includegraphics[scale=0.8]{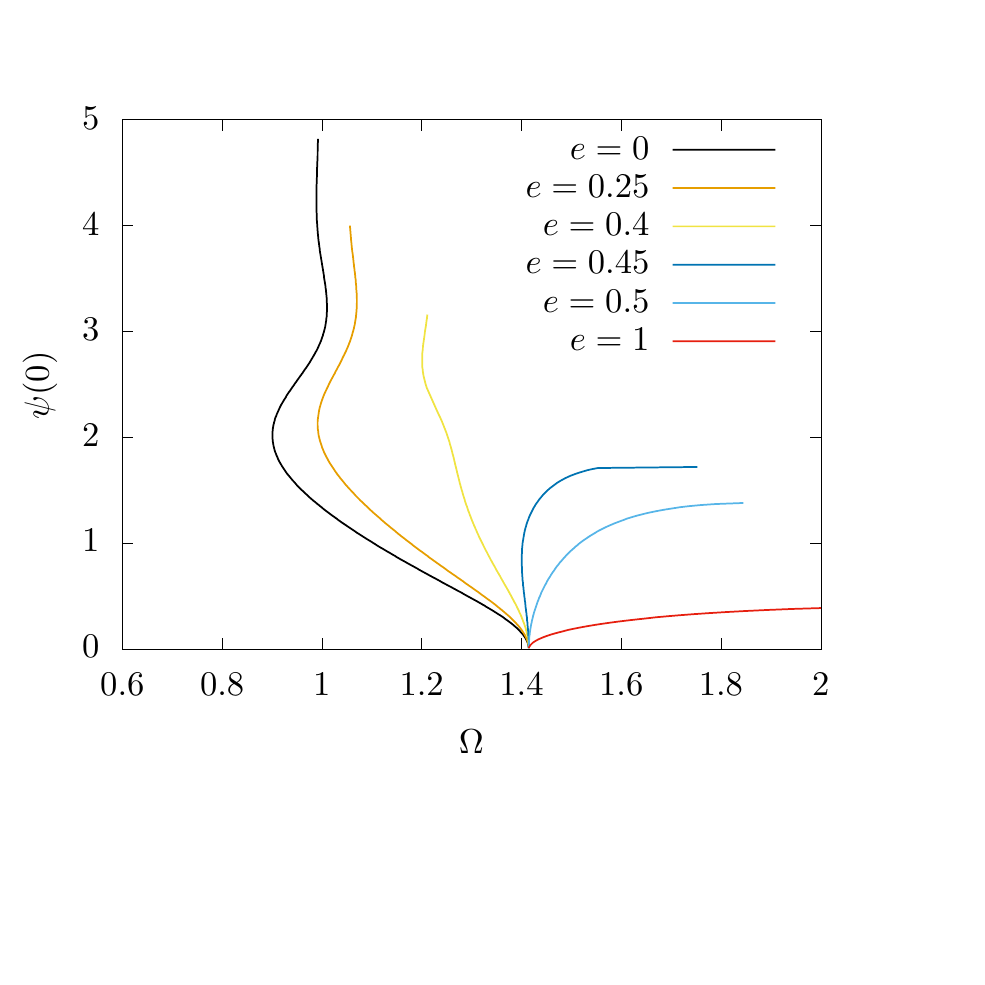}
    \includegraphics[scale=0.8]{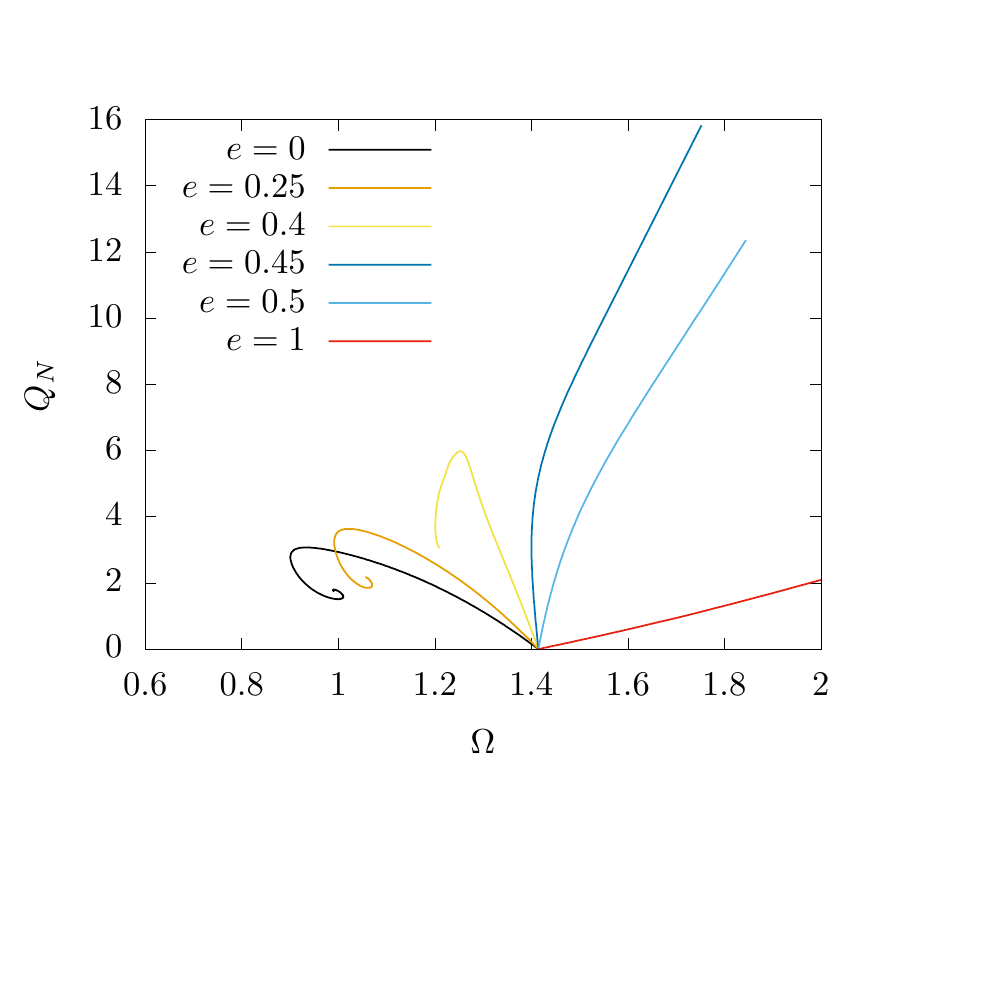}
    \vspace{-2cm}
    \caption{\emph{Left}: We show the value of the scalar field at the origin, $\psi(0)$, as a function of $\Omega$ for $\alpha = 0.1$, $\Lambda = -1/6$, $k=0$ and different values of $e$. \emph{Right}: The Noether charge $Q_N$ as function of $\Omega$ for the same solutions.}
    \label{fig:psi0_e_alp0_1} 
\end{figure}

To understand the influence of gravity, we give the values of $\psi_0$ and the Noether charge $Q_N$ in dependence
of $\Omega$ for $\alpha=0.1$ in Fig. \ref{fig:psi0_e_alp0_1}.  Comparing this with the results for the $\alpha=0$
case (see Fig. \ref{fig:psi0_e_alp0}) the interplay between the electromagnetic repulsion and gravitational
attraction becomes apparent.  For $\alpha=0.1$ the curve of $\psi(0)$ as function of $\Omega$ does not show the above mentioned back-bending behaviour until $e$ is large enough. This is also seen when considering the
Noether charge $Q_N$. For $e=0$, $e=0.25$ and $e=0.4$ the curves show the typical spiraling behaviour for
uncharged boson stars. Only when the electromagnetic interaction becomes sufficiently strong and dominates the system, does this behaviour change. We also observe when comparing with the $\alpha=0$ case that $\psi(0)$ is much larger
when backreaction of the space-time is not taken into account. This is related to the fact that when the space-time
backreacts a limit for the mass of any gravitating object exists~: $2M$ per Schwarschild radius.

\begin{figure}[h!]
\begin{center}
{\includegraphics[width=8cm]{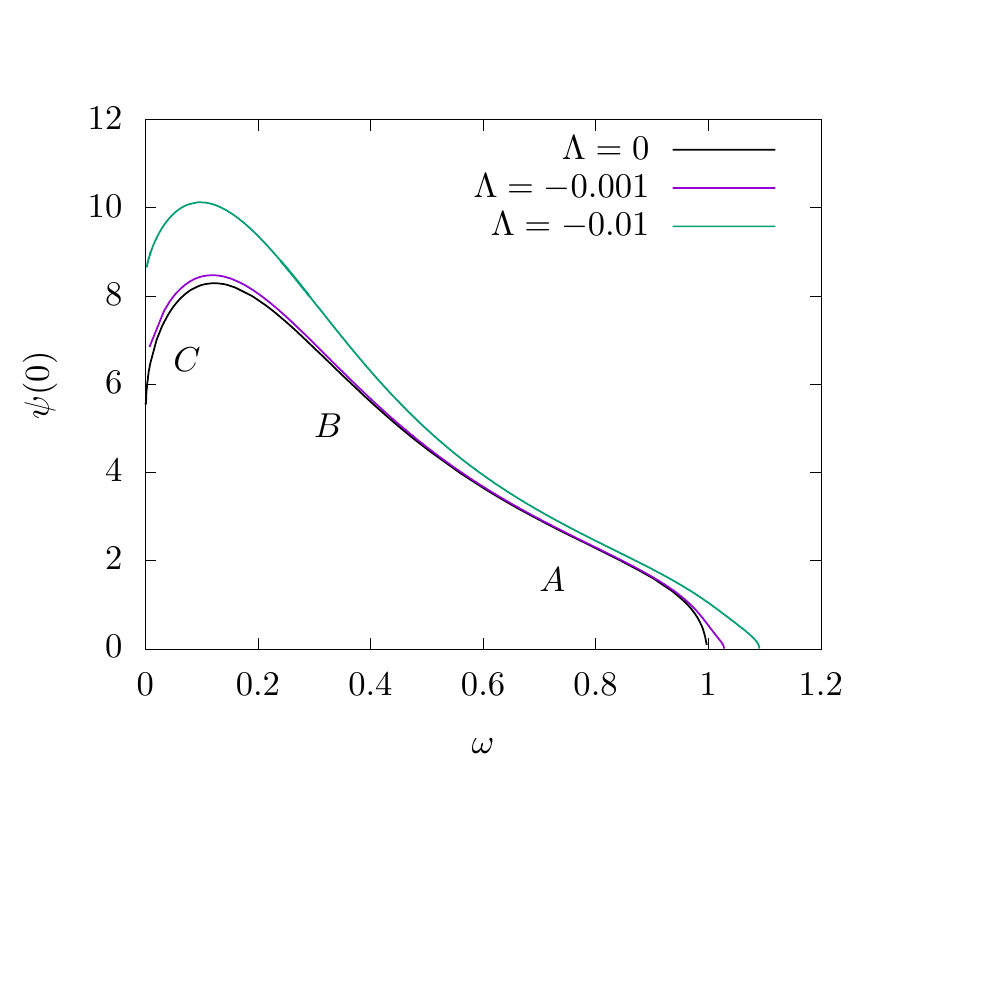}}
{\includegraphics[width=8cm]{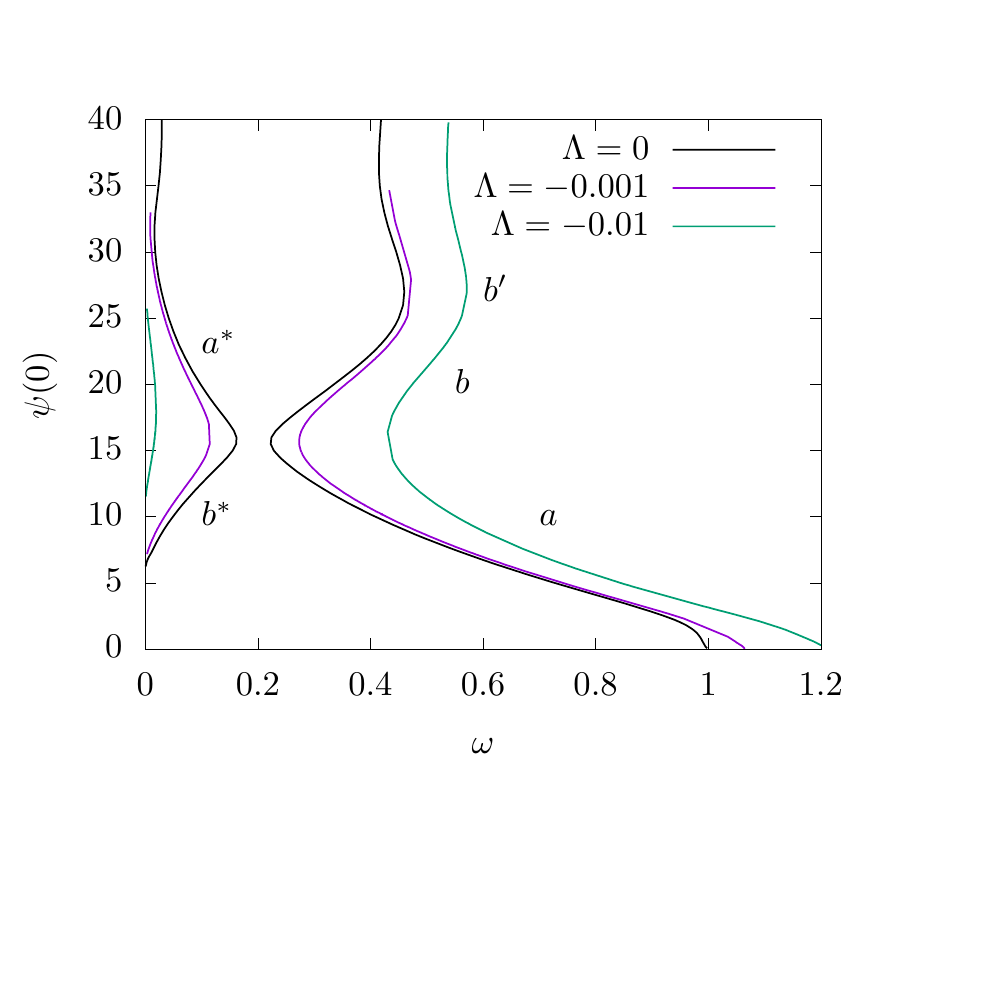}}
\vspace{-2cm}
\caption{{\it Left}: We show the dependence of $\psi(0)$ on $\omega$ for the unexcited solution ($k=0$) and three values of $\Lambda$ indicating branches $A$, $B$ and $C$. {\it Right}: We show the dependence of $\psi(0)$ on $\omega$ for the first radially excited solution ($k=1$) and three values of $\Lambda$ indicating branches $a$, $b$, $b^{\prime}$ and 
$a^*$, $b^*$, respectively, for three different values of $\Lambda$. For both cases we have chosen $\alpha=0.0012$ 
and $e=0.02$. \label{fig:data_1_N_extra} }
\end{center}
\end{figure}

\begin{figure}
    \centering
    \includegraphics[scale=1]{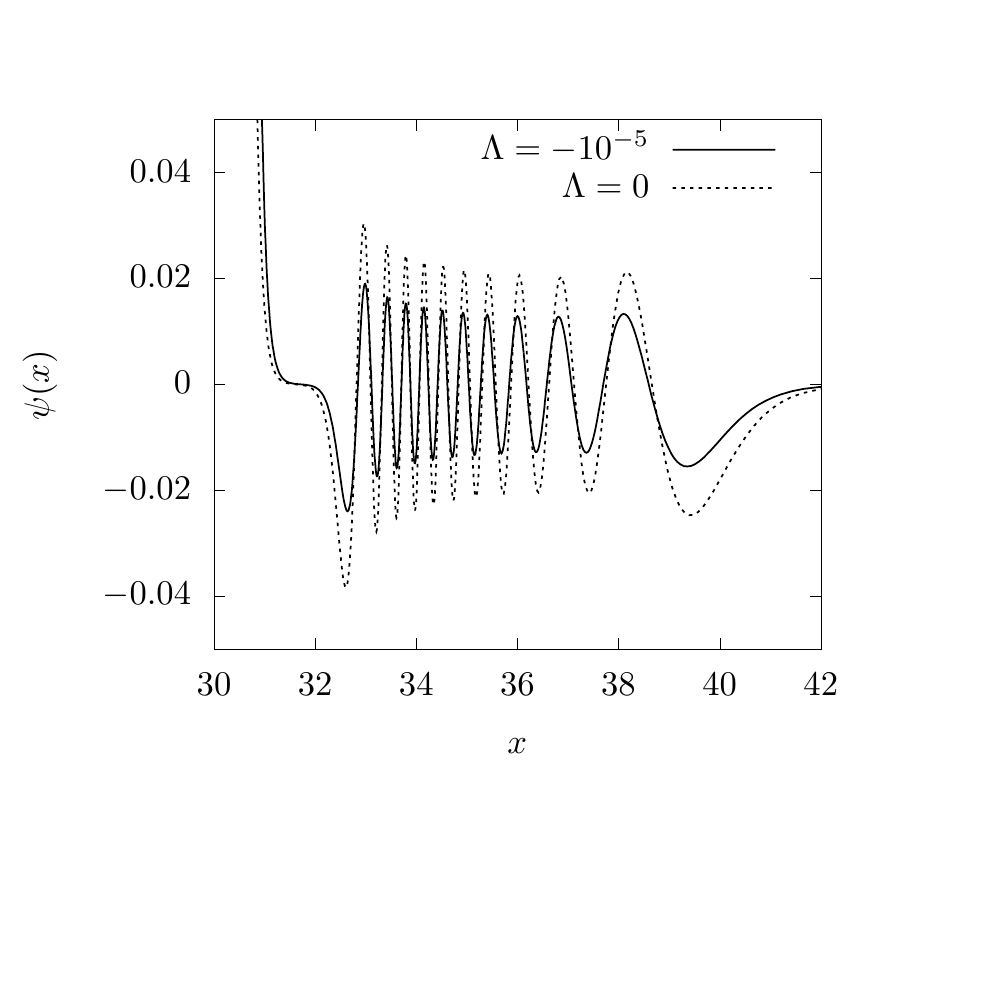}
    \vspace{-2.5cm}
    \caption{We show the profile of the scalar field function $\psi(x)$ for $\alpha = 0.0012$, $e=0.02$, $\omega = 0.001$ and $\Lambda=10^{-5}$ (solid) and $\Lambda=0$ (dashed), respectively. }
    \label{fig:wavy_scalar_lambda} 
\end{figure}

The branches $A$, $B$ and $C$ discussed in the asymptotically flat case exist as well in the $\Lambda < 0$ case,
see Fig. \ref{fig:data_1_N_extra} (left), where we give $\psi(0)$ in function of $\omega$ for three different
values of $\Lambda$ including the already discussed asymptotically flat case. Clearly, the branches $A$, $B$ and $C$ are present for $\Lambda < 0$ and the qualitative dependence of $\psi(0)$ on $\omega$ does not change significantly when changing $\Lambda$. In fact, we find that the wavy scalar hair solutions discussed above also exist in the
$\Lambda < 0$ case. In Fig. \ref{fig:wavy_scalar_lambda} we show the scalar field function for $\alpha=0.0012$, $e=0.02$, $\omega=0.001$ and $\Lambda=10^{-5}$ and compare it with the corresponding solution for $\Lambda=0$ (see Fig. \ref{fig:profile1} (upper right)). We observe that when changing $\Lambda$ that the first three local maxima are slightly shifted to larger $x$, while the seven remaining maxima are slightly shifted to smaller values of $x$, i.e. the oscillations appear in a smaller interval of $x$. More significantly, however, we observe also that
the amplitude of the oscillations decreases when decreasing $\Lambda$ from zero. We would expect that decreasing $\Lambda$ further, i.e. decreasing the AdS radius, will suppress the oscillations as soon as the AdS radius becomes comparable to the $x$-values at which these oscillations appear.

\begin{figure}
    \centering
    \includegraphics[scale=0.8]{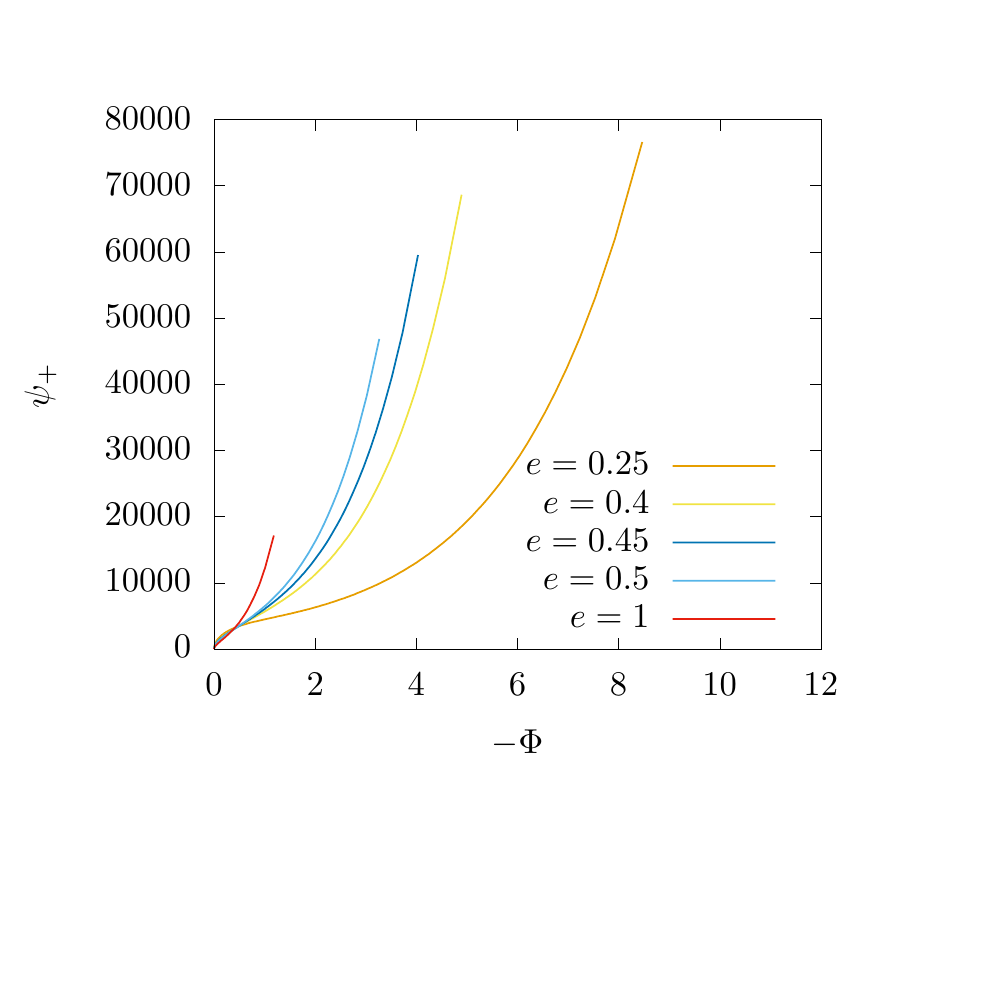}
    \includegraphics[scale=0.8]{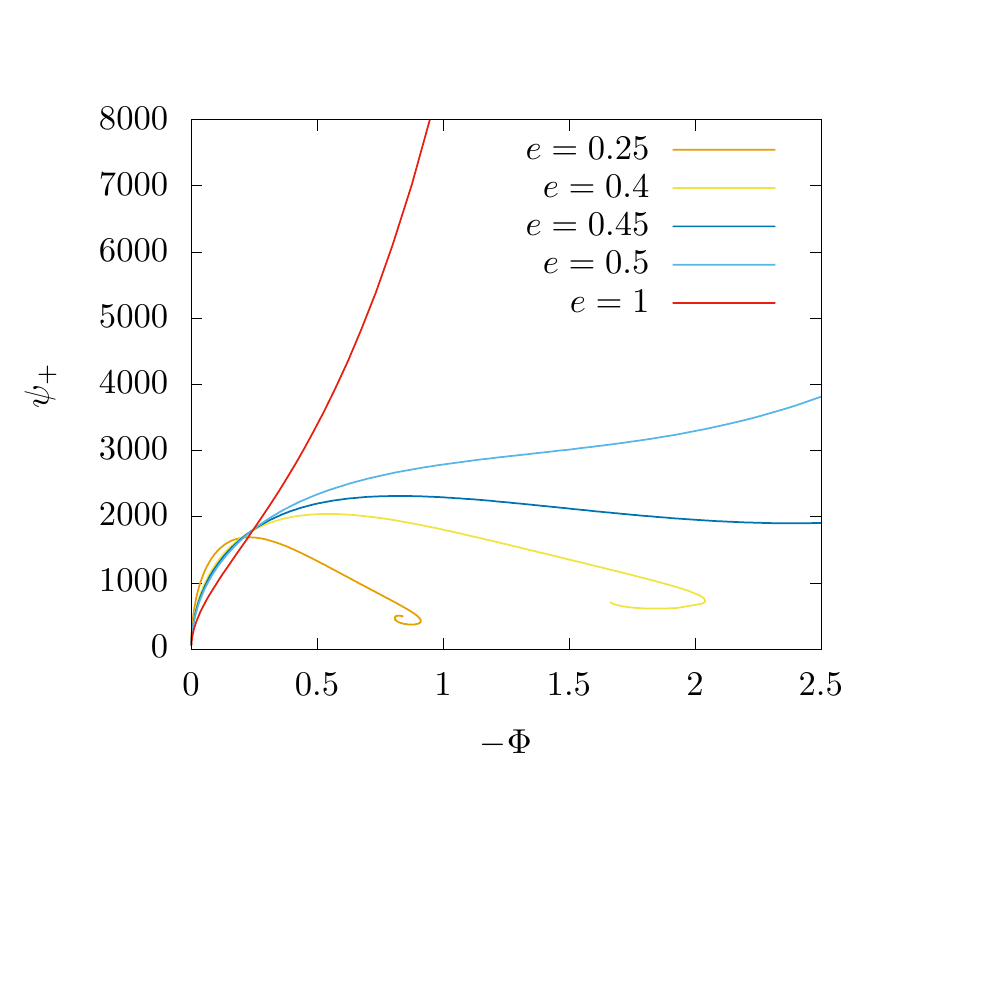}
    \vspace{-2cm}
    \caption{\emph{Left}: The value of the condensate, $\psi_+$, as a function of the chemical potential, $-\Phi$, for several values of $e$, $\alpha = 0.01$ and $\Lambda = -1/6$. \emph{Right}: Same as left, but for  $\alpha = 0.1$.}
    \label{fig:psi_+_vs_Omega_alp_0_1}
\end{figure}

In order to make a connection to the AdS/CFT correspondence and the suggested interpretation of the field values of solitonic
solutions in aAdS on the conformal boundary in terms of glueball condensates (see e.g. \cite{Horowitz:2010jq, Hartmann_Riedel}), we
plot the value of $\psi_+$ as function of $-\Phi$ for $\Lambda = -1/6$, several
values of $e$ and $\alpha=0.01$ (left) and $\alpha=0.1$ (right) in Fig. \ref{fig:psi_+_vs_Omega_alp_0_1}. This is hence a holographic phase diagram showing the condensate in function of the chemical potential. Increasing the backreaction of the space-time, i.e. increasing $\alpha$, changes the qualitative behaviour of the condensate. At given fixed value of the gauge coupling $e$, the increase of the potential difference between the origin and the conformal boundary $V(0)-V(r\rightarrow\infty)=-\Phi$ increases the value of the condensate up to a maximal value of both the potential difference and the condensate value. We observe that this is true for small $\alpha$ and $e$ and find that the maximal possible value of the condensate decreases with increasing $e$ and $\alpha$. For larger values of $\alpha$ and sufficiently large values of $e$ we observe that at a given value of $-\Phi$ the condensate reaches a maximum and then decreases again when increasing $-\Phi$ further. In fact, we find that for $e$ large enough a spiraling behaviour of the condensate value appears, i.e. we observe that for a given values of $e$ and $\alpha$ an interval in $-\Phi$ exists on which different values of the condensate are possible for the same value of $-\Phi$.

\subsection{Radially excited solutions}

Following our discussion above, we would expect that radially excited boson stars in aAdS should exist in our model. 
In fact, they also exist for the $\Lambda=0$ limit and since these solutions have not been
discussed in the literature so far, we will first present our results for the asymptotically flat case.

\subsubsection{$\Lambda=0$}

In Fig. \ref{fig:flat_unexcited} we show the mass $M$ and the Noether charge $Q_N$ of the radially excited boson stars with one node in the scalar field function ($k=1$) for $\alpha=0.0012$ and $e=0.02$ in dependence on $\Omega$. We find that the unexcited and the one-node solutions show a qualitatively similar behaviour 
in the limit $\psi(0) \to 0$, e.g. $\Omega\rightarrow 1$ for both $k=0$ and $k=1$ and
the scalar field function tends uniformly to zero while the mass $M$ and Noether charge $Q_N$ tend to non-vanishing values.

\begin{figure}
    \centering
    \includegraphics[scale=0.8]{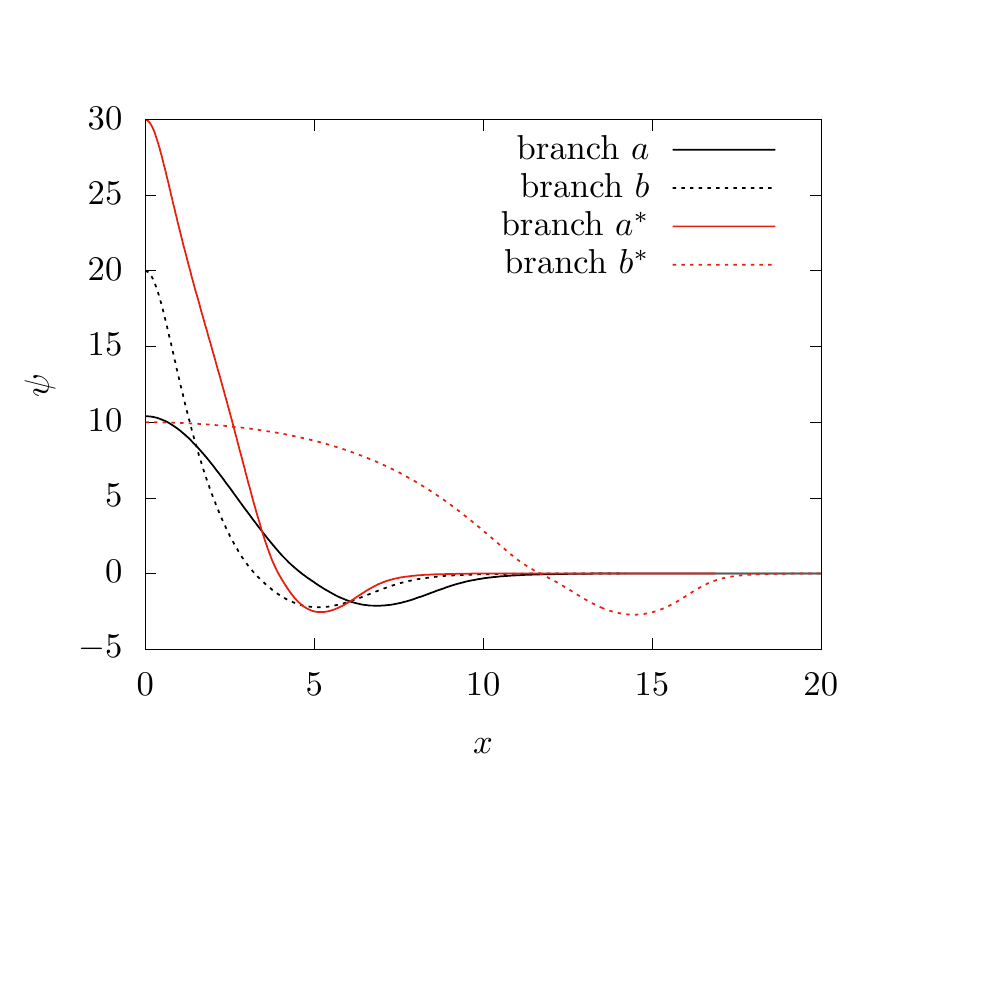}
    \includegraphics[scale=0.8]{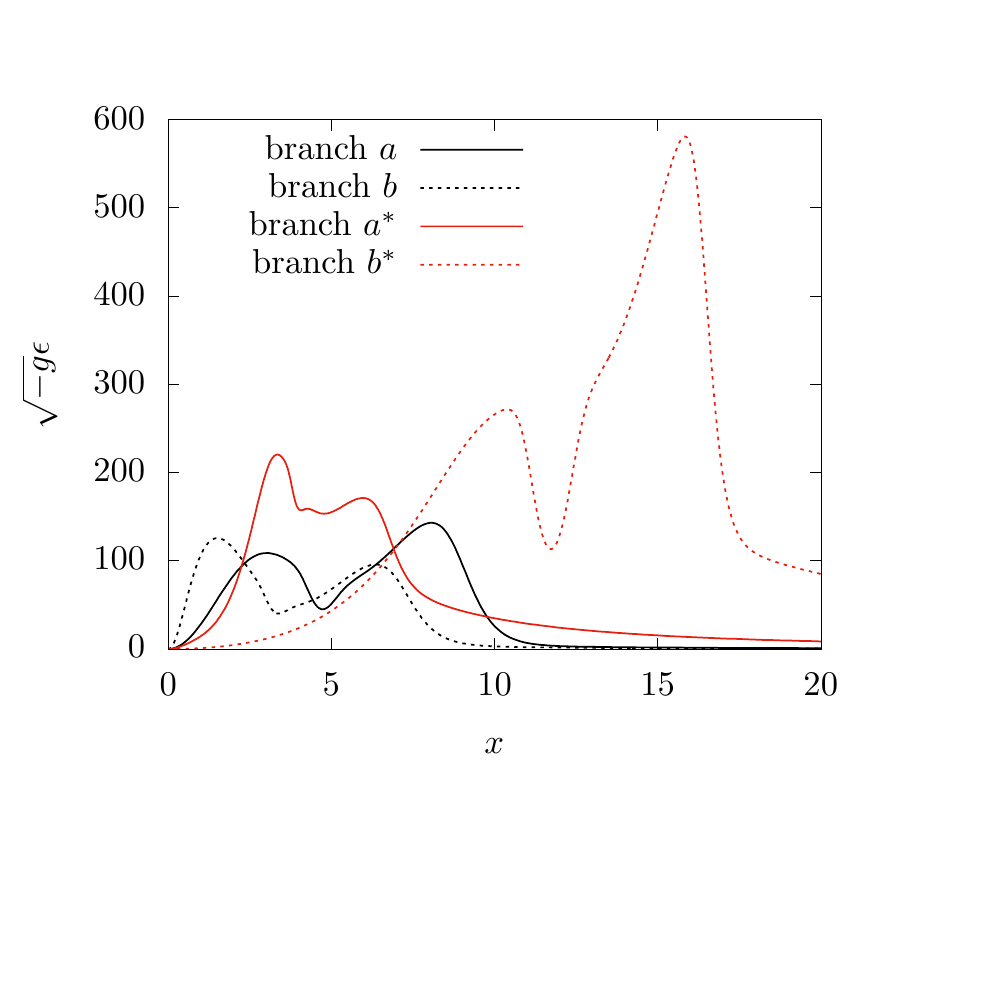}\\
    \vspace{-2cm}
    \includegraphics[scale=0.8]{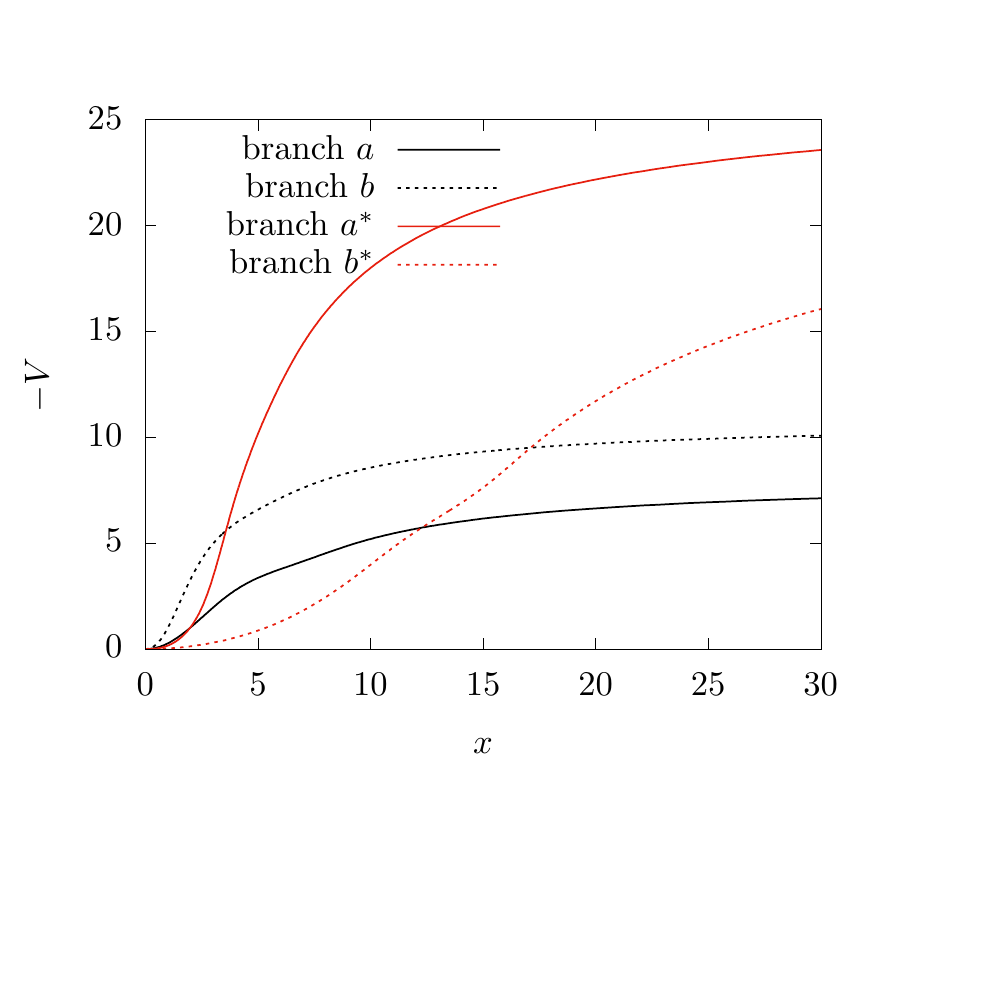}
    \includegraphics[scale=0.8]{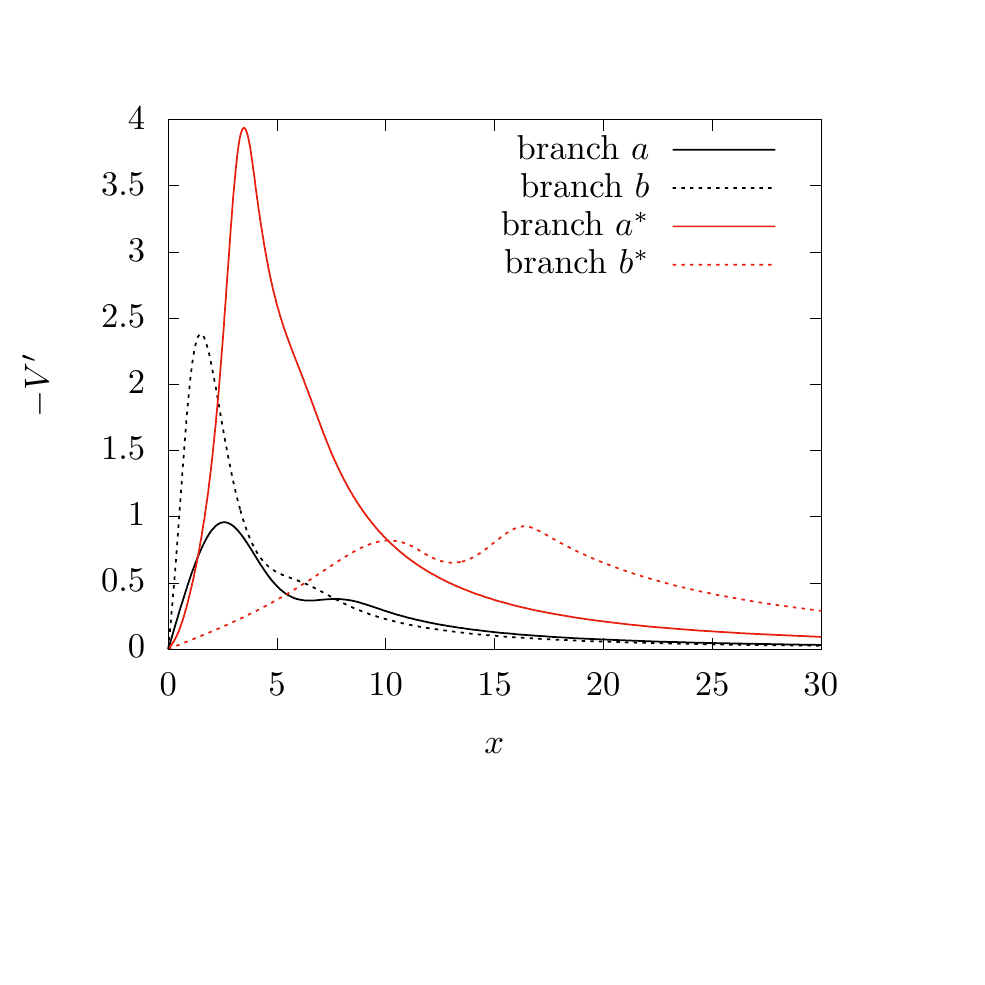}\\
    \vspace{-2cm}
    \includegraphics[scale=0.8]{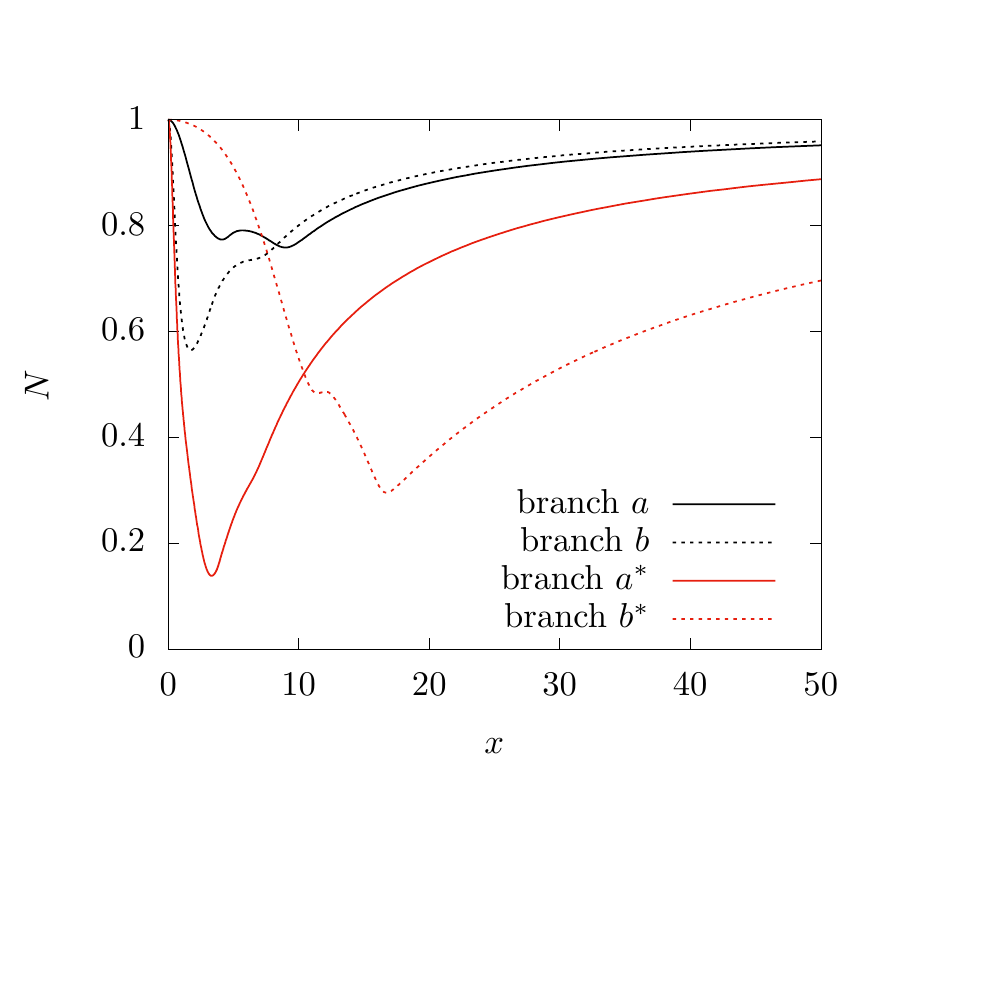}
    \includegraphics[scale=0.8]{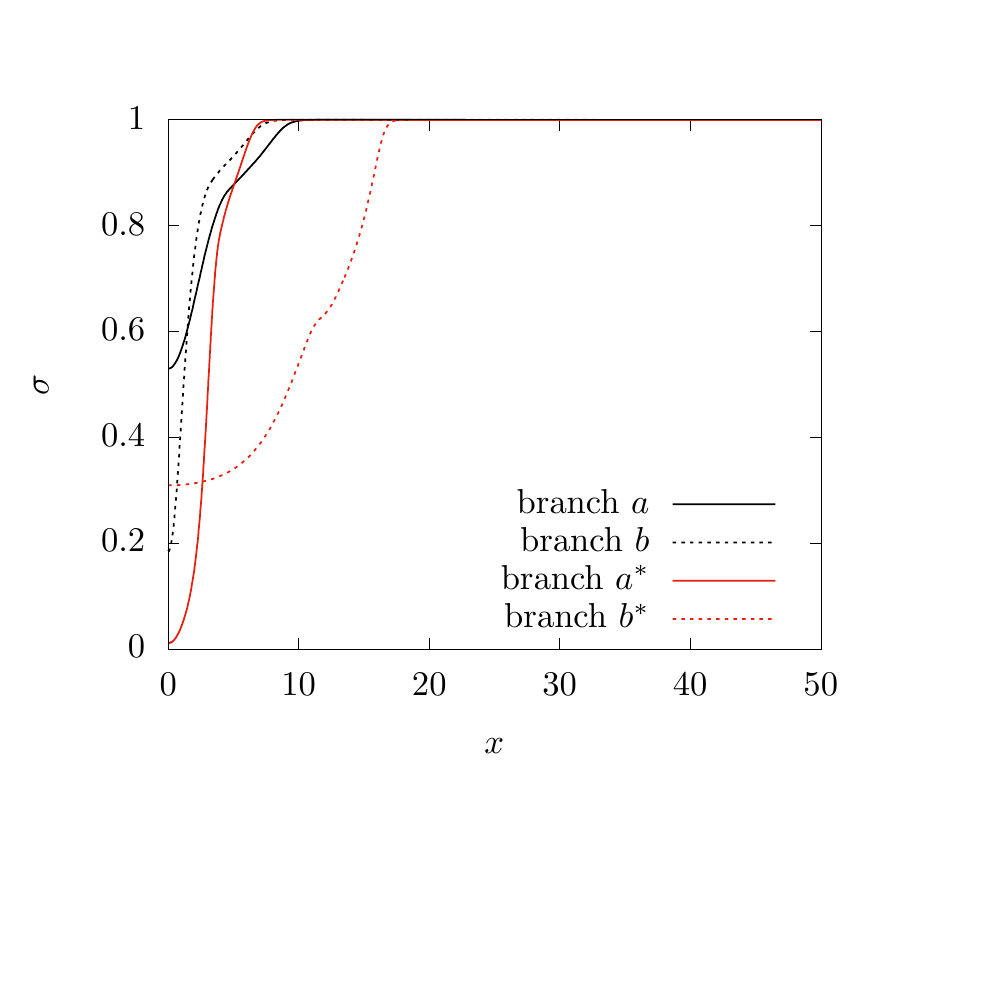}
    \vspace{-2cm}
    \caption{We compare the four radially excited solutions on branches $a$, $b$, $a^*$ and $b^*$ for $k=1$, $\alpha =0.0012$, $e=0.02$, $\Lambda=0$ and $\Omega=0.53$, see Fig.\ref{fig:data_1_N_extra}.
    {\it Top Left}: Scalar field function $\psi(x)$, {\it Top Right}: effective energy density $\sqrt{-g}T_0^0=\sqrt{-g}\epsilon$. 
    {\it Middle Left}: Electric potential $-V(x)$, {\it Middle Right}: Electric field $-V^{\prime}(x)$, 
    {\it Bottom Left}: Metric function $N(x)$, {\it Bottom Right}: Metric function $\sigma(x)$.}
    \label{fig:all_branches_k1}
\end{figure}

The behaviour, however, changes when increasing $\psi(0)$. While we also find several branches for the radially excited case, these have very different features as compared to the branches discussed for $k=0$. Rather than finding branches $A$, $B$ and $C$, the excited solutions form several branches in the form of a spiral for which we will use the labels $a$, $b$, $b^{\prime}$ (for the first three branches) in the following.
In fact, the second branch of excited solutions has lower  mass and Noether charge, respectively, as the first, major branch that connects to $\Omega=1$. To understand this result, we show the scalar
field and the associated effective energy density $\sqrt{-g} T^0_0=\sqrt{-g} \epsilon$ in  Fig.\ref{fig:all_branches_k1} (upper row) for $\Omega=0.55$ and the branches $a$ and $b$. 
We observe that the solutions on branch $b$ have larger central value of the scalar field, $\psi(0)$, but that the scalar field's node is at smaller $x$.  Moreover, we find that the effective energy density possesses two local maxima, one close to the node of the scalar field function. These maxima are more pronounced for solutions on branch $a$.

In Fig. \ref{fig:all_branches_k1} (middle row) we show the electric potential $V(x)$ and the electric field
$-V'(x)$ for the same solutions. Both potential and electric field are stronger close to the core of the boson star for the solutions on branch $b$ as
compared to those on branch $a$. Consequently, the local space-time curvature is also stronger there, see 
the metric functions $N(x)$ and $\sigma(x)$ in Fig. \ref{fig:all_branches_k1} (bottom row).

\begin{figure}
    \centering
    \includegraphics[scale=0.8]{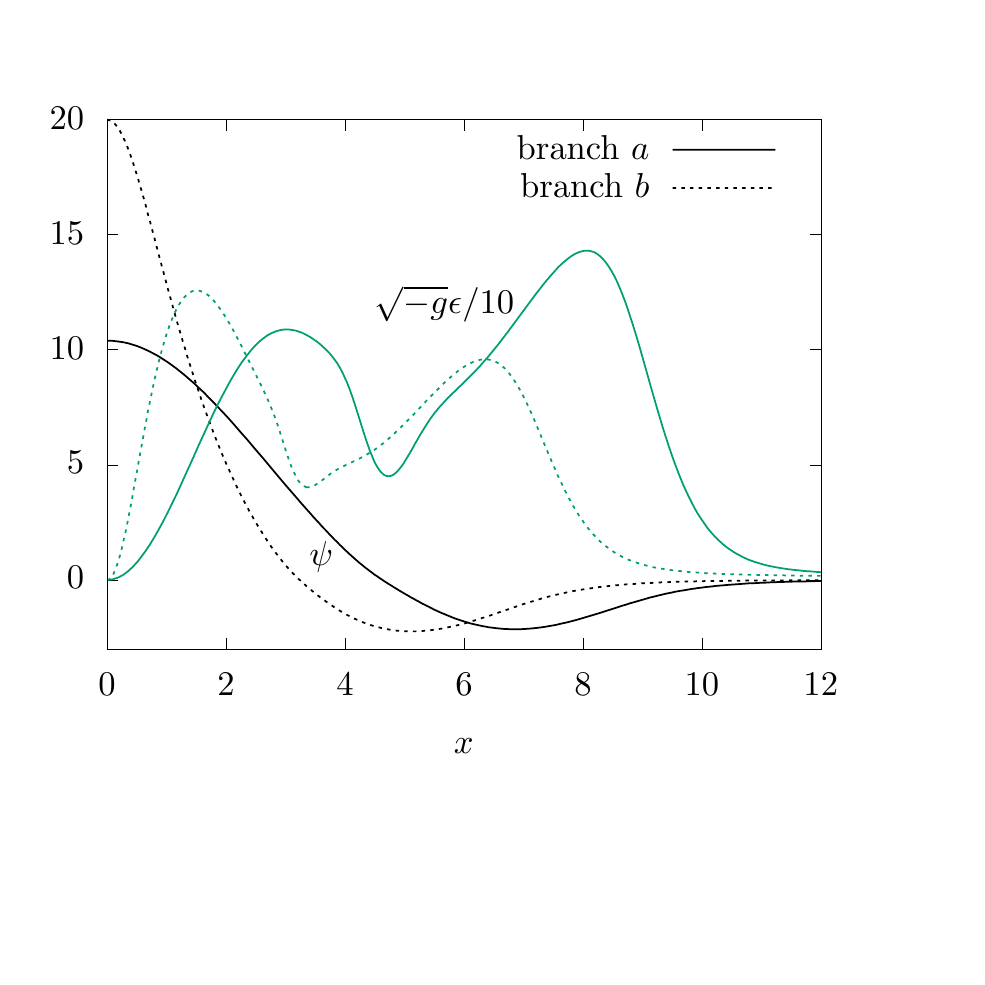}
    \includegraphics[scale=0.8]{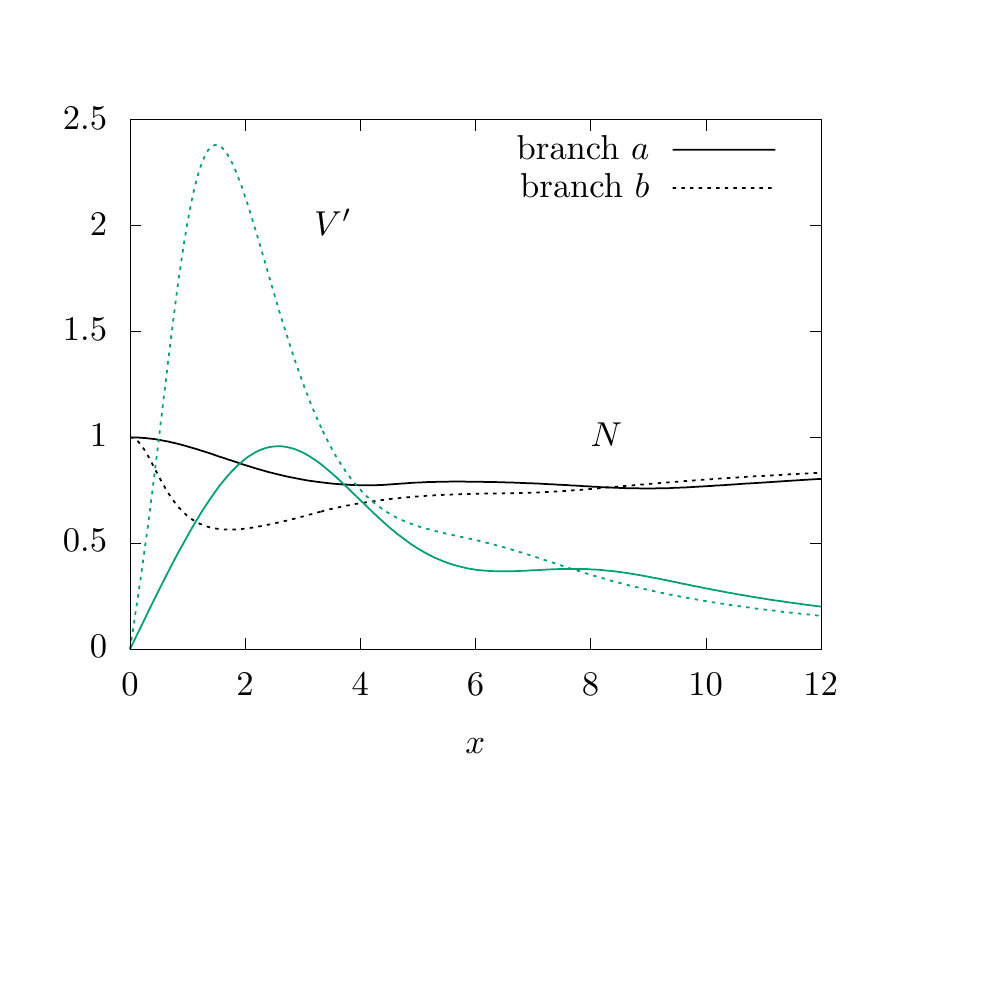}
    \vspace{-2cm}
    \caption{We compare the two radially excited solutions on branch $a$ and branch $b$ for $k=1$, $\alpha =0.0012$, $e=0.02$, $\Lambda=0$ and $\Omega=0.55$, see Fig.\ref{fig:flat_unexcited} (right). 
    {\it Left}: Scalar field function $\psi(x)$ and effective energy density $\sqrt{-g}T_0^0=\sqrt{-g}\epsilon$. 
    {\it Right}: Metric function $N(x)$ and negative of electric field $V^{\prime}(x)$. }
    \label{fig:profile_1_N}
\end{figure}

Interestingly, we find that next to the branch of solutions discussed above  another, completely disconnected branch exists that
appears only in the highly non-linear regime. 
This is shown in Fig. \ref{fig:data_1_N_extra} (right) where we give the value of $\psi(0)$ in dependence
on $\omega$. To compare, we have also plotted again the values for the branches $a$, $b$ and $b^{\prime}$.
At sufficiently small $\omega$
and sufficiently large $\psi(0)$ we find two new branches $a^*$ and $b^*$ that are disconnected from the other branches and do not connect to the $\omega=1$, $\psi(0)=0$ limit. In keeping with the notation used above, the
$b^*$ part of these new solutions has higher mass as compared to $a^*$ part.

In order to understand the difference between these one-node solutions, we show the scalar field function $\psi(x)$ as well as the effective energy density $\sqrt{-g} \epsilon$ of the four $k=1$ solutions for $\Omega = 0.53$ in Fig. \ref{fig:all_branches_k1} (upper row),  the corresponding electric potential $V(x)$ and field $-V'(x)$, respectively, are shown in Fig. \ref{fig:all_branches_k1} (middle row), while we give the metric functions $N(x)$ and $\sigma(x)$, respectively, in Fig.\ref{fig:all_branches_k1} (bottom row). 
Interestingly, branches $a$, $b$ and $a^*$ have the zero of the scalar field function roughly at the same value of $x$, while the zero of the solution on branch $b^*$ is at much larger values of $x$. The corresponding maximum of the effective energy density $\sqrt{-g}\epsilon$ as well as the minimum of the metric function $N(x)$ consequently also appear at larger $x$ for branch $b^*$ as compared to the remaining branches.

\subsubsection{$\Lambda<0$}

\begin{figure}
\centering
\includegraphics[scale=0.8]{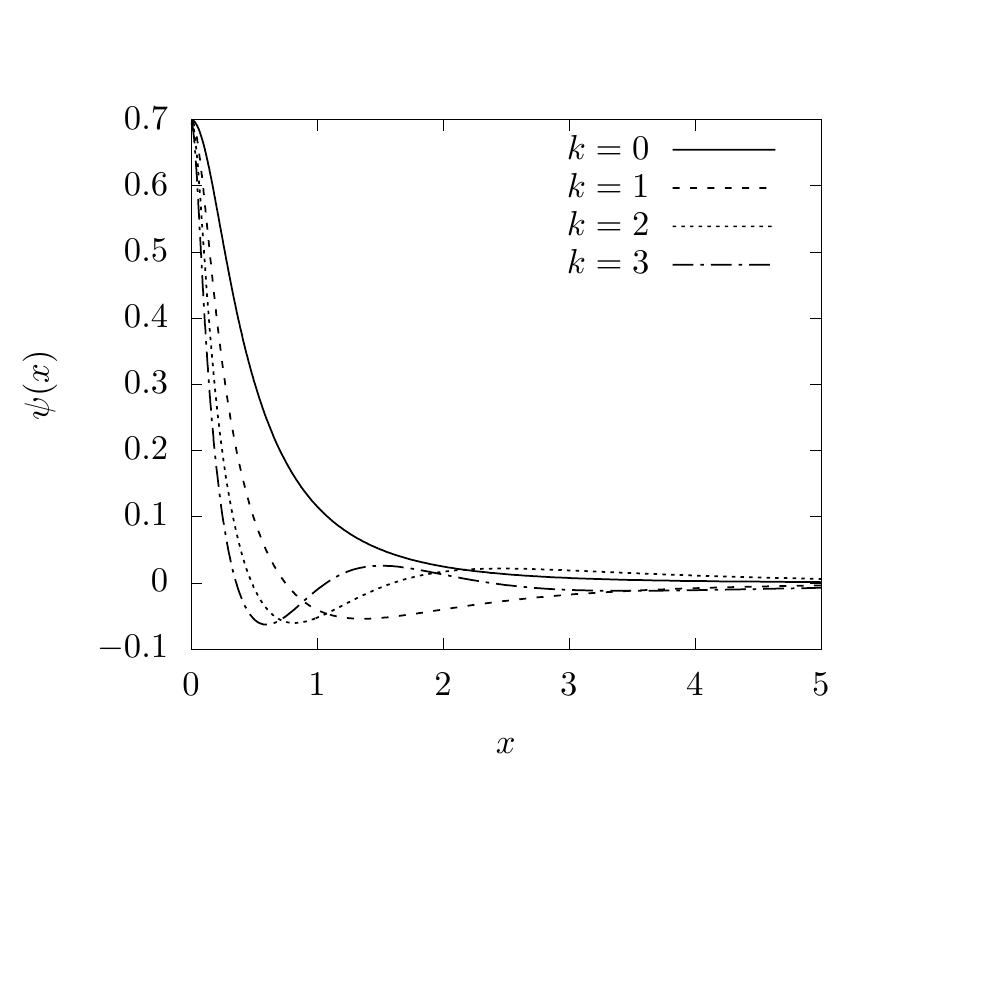} \includegraphics[scale=0.8]{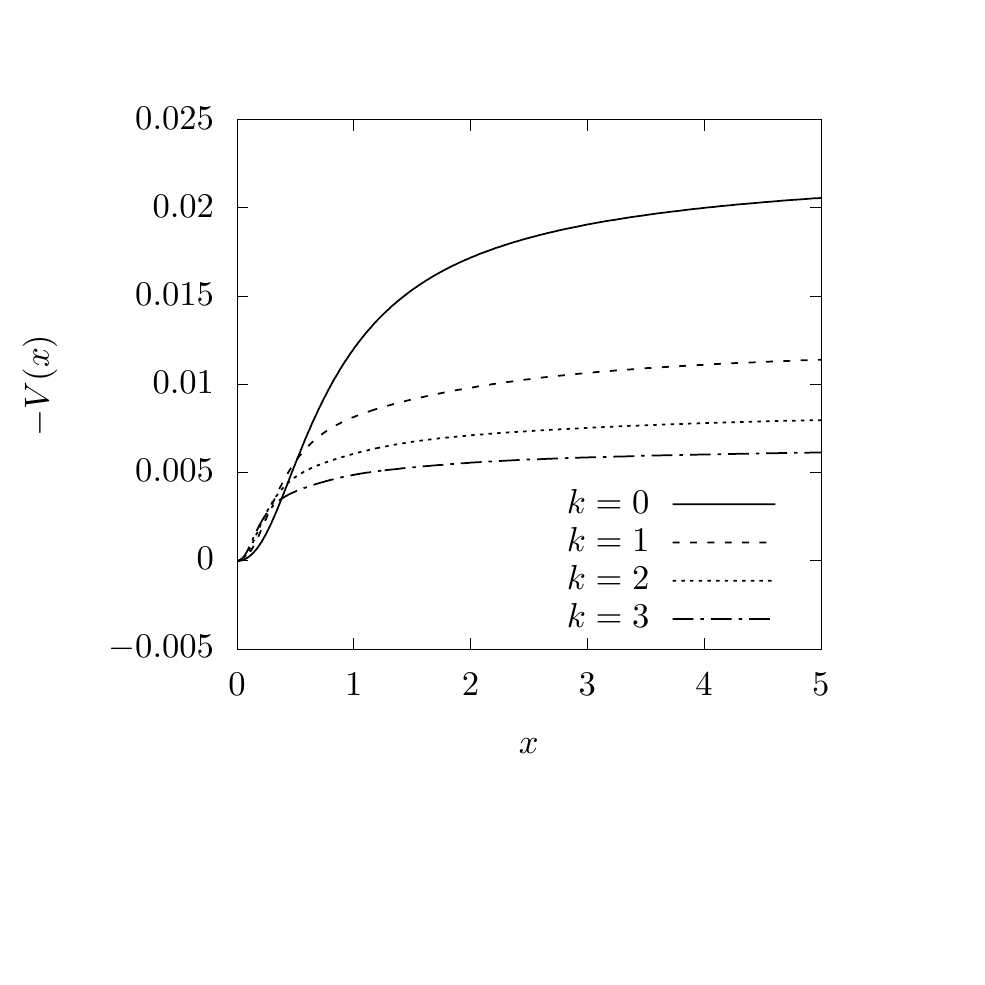}
\vspace{-2cm}
\caption{(\emph{Left}): We show the scalar field function $\psi(x)$ of the unexcited solution ($k=0)$ and the first three radially excited solutions ($k=1$, $2$, $3$) for $\alpha = 1$, $e = 0.1$ and $\Lambda = -3/4$. (\emph{Right}): The corresponding gauge field function $-V(x)$. }
 \label{fig:psi_and_V}
\end{figure}

\begin{figure}
\centering
\includegraphics[scale=0.8]{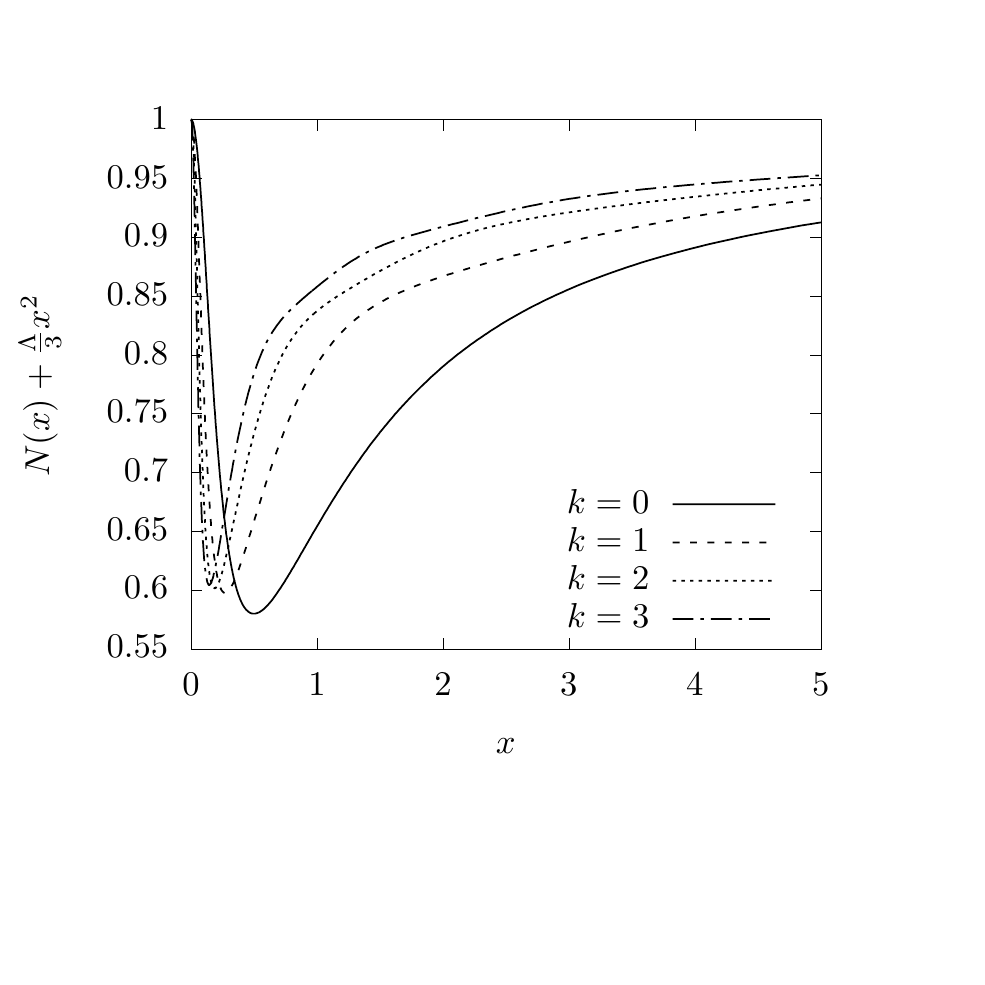} \includegraphics[scale=0.8]{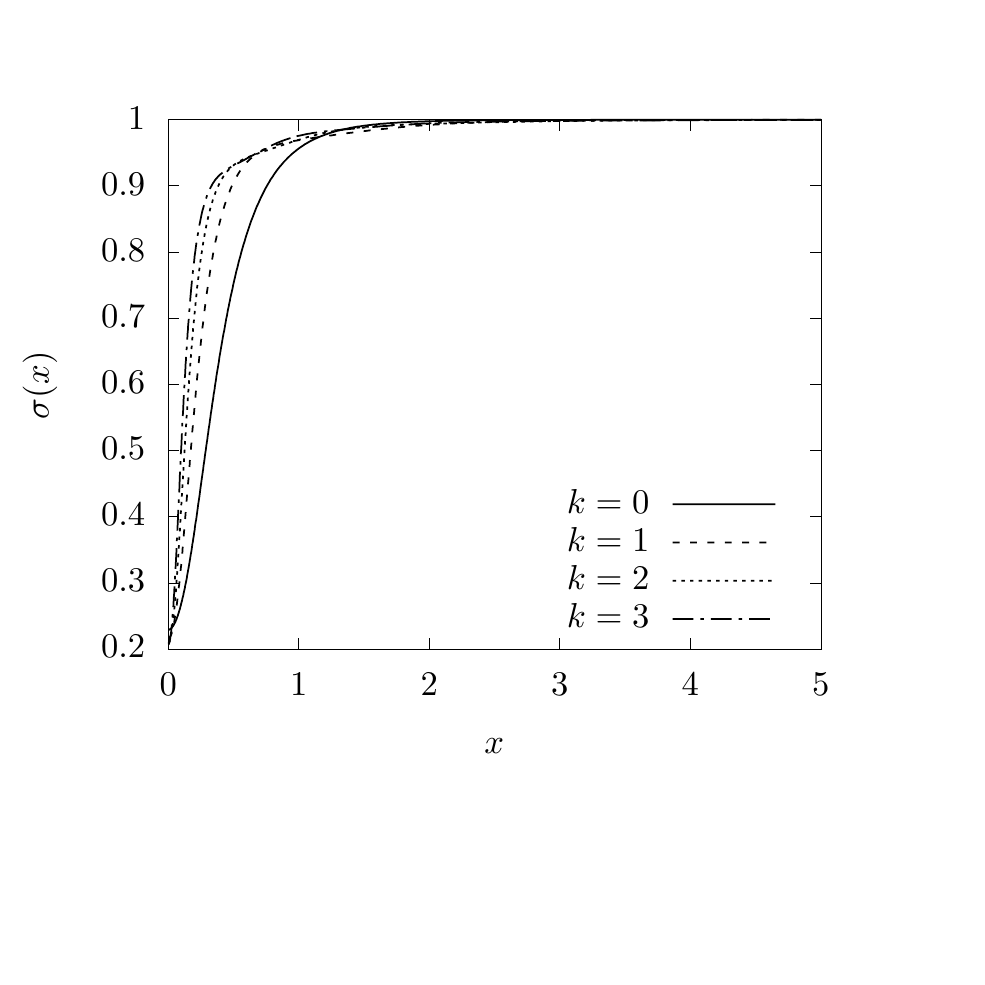}
\vspace{-2cm}
\caption{ 
We show the metric function $N(x)+\frac{\Lambda}{3}x^2$ of the unexcited solution ($k=0)$ and the first three radially excited solutions ($k=1$, $2$, $3$) for $\alpha = 1$, $e = 0.1$ and $\Lambda = -3/4$. (\emph{Right}): The corresponding metric function $\sigma(x)$.}
\label{fig:n_and_sigma}
\end{figure}

We have constructed these solutions numerically. In Fig. {\ref{fig:psi_and_V}} and Fig. \ref{fig:n_and_sigma} we show the matter and metric functions, respectively, of solutions with $k=1,2,3$ zeros in the scalar field function corresponding to $\Omega \approx 2.32, \, 3.23, \, 4.15$, respectively, for $\alpha = 1$, $e =0.1$ and $\Lambda = -3/4$. Increasing $k$, we observe that the minimum of the metric function $N(x)$ decreases in value and moves closer to the origin, while $\sigma(x)$ reaches its asymptotic value quicker when increasing $k$.

\begin{figure}
\centering
\includegraphics[scale=0.8]{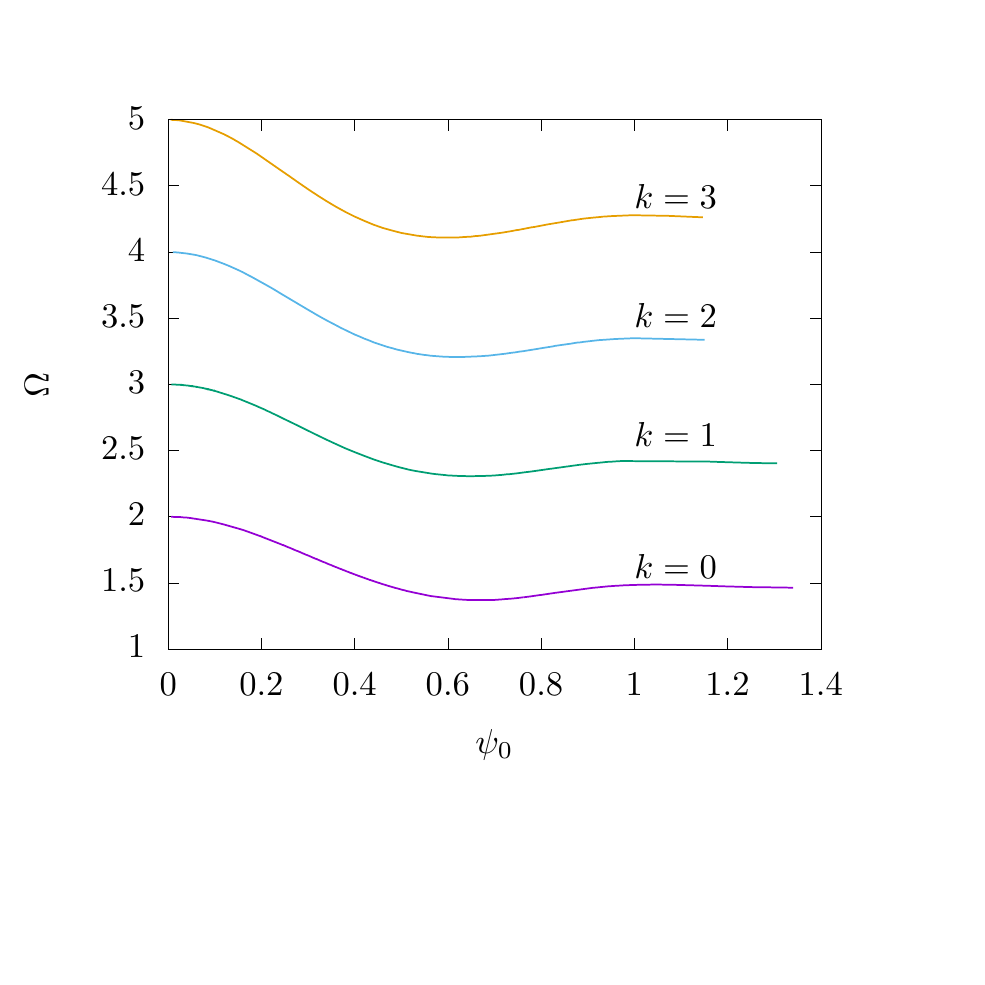} \includegraphics[scale=0.8]{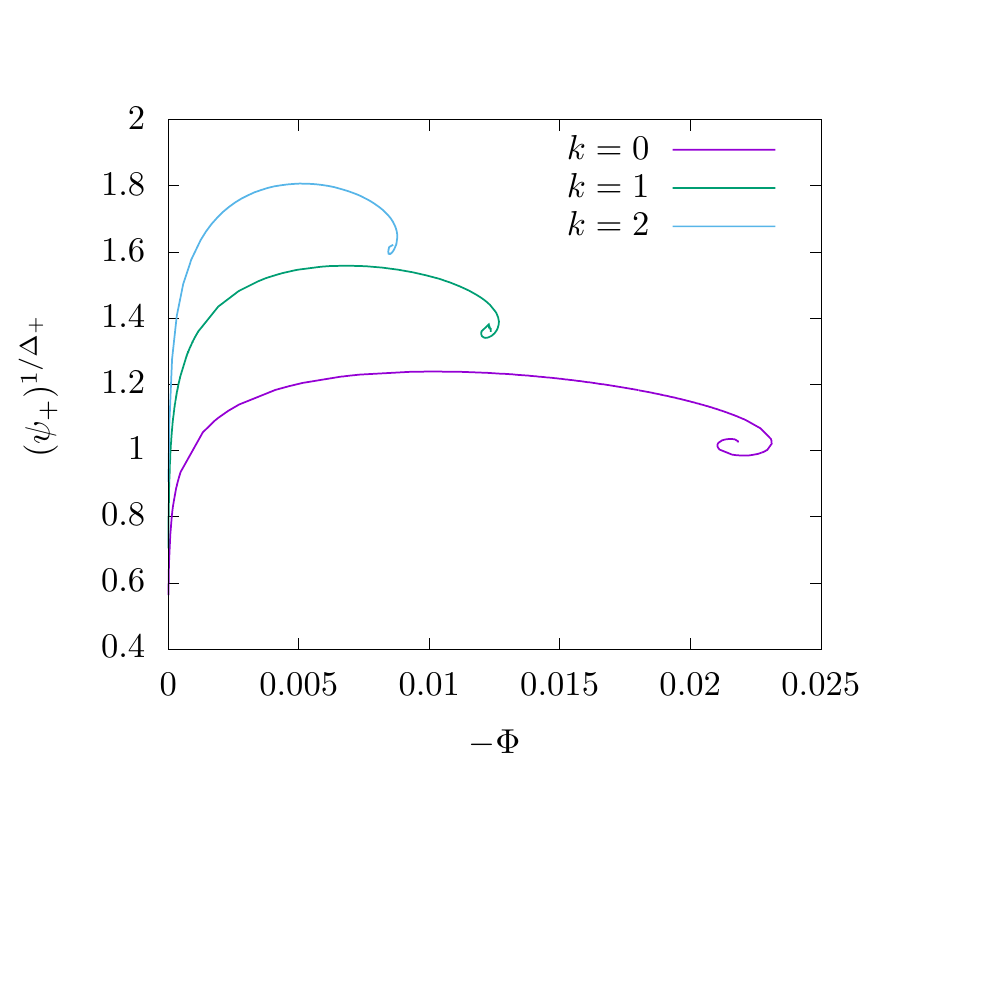}
\vspace{-2cm}
\caption{(\emph{Left}): We show the value of $\Omega$ in dependence of the value of the scalar field function at the origin, $\psi_0$, of the unexcited solution ($k=0)$ and the first three radially excited solutions ($k=1$, $2$, $3$) for $\alpha = 1$, $e = 0.1$ and $\Lambda = -3/4$. (\emph{Right}): The corresponding condensate $(\psi_+)^{1/\Delta_+}$ as function of the chemical potential $-\Phi$ for $k=0,1,2$. \label{fig:omega_condensate}}
\end{figure}

In Fig. \ref{fig:omega_condensate} (left) we give $\Omega$ as function of the scalar field function at the origin, $\psi(0)\equiv \psi_0$ for solutions with $k=0,1,2,3$ nodes in the scalar field
for $\mu=1$, $\Lambda=-3/4$. Using the discussion on oscillons (see \ref{subsection:oscillons}) we have hence $p_+=2$ and consequently $\Omega = 2+k$, $k\in \mathbb{N}$.
The data in Fig. \ref{fig:omega_condensate} clearly demonstrates that 
for small values of $\psi_0$ we find that $\Omega = 2+k$, where $k$ is the number of nodes of the scalar field function in agreement with the discussion given above. Increasing $\psi_0$ leads to a decrease in the value of $\Omega$ until $\Omega_{\rm min}$ is reached at some intermediate value of $\psi_0$. We find that $\Omega$ reaches its minimal value at decreasing values of $\psi_0$ when increasing the number of nodes in the scalar field. 

In Fig. \ref{fig:omega_condensate} (right) we give the value of the condensate, i.e. the value of the scalar field on the conformal boundary, $(\psi_+)^{1/\Delta_+}$, in function of the chemical potential $-\Phi$. For all $k$ we find that the condensate
builds up strongly when increasing $-\Phi$ from zero. It then reaches a maximal value and this is attained 
at decreasing values of $-\Phi$ when increasing $k$. The condensate then decreases up to a maximal value of 
$-\Phi$ and starts to form a spiral with several branches appearing. We observe a similar behaviour for all
$k$, but note that the interval in $-\Phi$ for which solutions exist decreases with $k$.
Interestingly, we also find that the value of the condensate increases when increasing $k$ at a fixed chemical potential and that different
values of the condensate are possible for sufficiently large and fixed $-\Phi$.

Interestingly, we find that the new  branches of solutions found in the $\Lambda=0$ limit exist as well for $\Lambda <0$, at least as long as $\Lambda$ is sufficiently close to zero. This is shown in Fig. \ref{fig:data_1_N_extra} (right). The branches $a$, $b$ and $b^{\prime}$ move to larger values of $\omega$, while the branches $a^*$ and $b^*$
move to smaller values of $\omega$. For $\Lambda = -0.01$, the $a^*$ and $b^*$ branches have nearly disappeared.
Hence, these branches exist only for $\Lambda$ sufficiently close to zero.

\section{Conclusions}
In this paper, we have studied electrically charged boson stars with an exponential scalar field interaction in asymptotically flat and Anti-de Sitter space-time, respectively.

Next to the standard boson stars discussed in the literature previously, we find several new branches of solutions.
For the unexcited solutions ($k=0$), i.e. solutions without nodes in the scalar field function
inside the boson star core, we observe that the formation of wavy scalar hair  on the boson stars discussed before in asymptotically flat space-time \cite{gauged_BS_power_law} also exists in aAdS space-time, at least for small values of $\vert \Lambda\vert$. This hair appears for gravitational interaction sufficiently large and electromagnetic interaction sufficiently small and involves the splitting of the space-time into a ''false vacuum'' interior (where the scalar field is constant, but non-zero) and a ''true vacuum'' exterior (where $\psi\equiv 0$). The interior hence corresponds to a de Sitter space-time with the constant scalar field leading to a constant positive potential energy that can be interpreted as a positive cosmological constant. The exterior corresponds to a space-time without 
scalar field, i.e. an extremal Reissner-Nordstr\"om (-AdS) solution. At the intersection between these space-times, the scalar field starts oscillating.  

When considering radially excited solutions, we find that the wavy scalar hair is not present, which also confirms that the spatial oscillations interpreted as wavy scalar hair are {\bf not} radial excitations of the boson star. In fact, we observe a qualitatively very different behaviour. When increasing the central value of the scalar field, instead of reaching a maximal possible value as in the $k=0$ case, we find that the central value can become much larger and seems (at least within our numerical analysis) not to be bounded from above as soon as the boson star is
radially excited. Moreover, we have constructed new branches of radially excited solutions that seem disconnected from the linear scalar field limit and seem to be a truly non-linear phenomenon. These branches disappear when decreasing $\Lambda$ to strongly from zero, i.e. exist only when the AdS radius is sufficiently large.

Finally, within the context of the holographic interpretation, we find that the radially excited solutions
can produce larger values of the condensate, i.e. the scalar field value on the conformal boundary, as compared to the unexcited solutions. In order to get a better understanding of these condensates, it will be interesting in the future to construct the corresponding black hole solutions and find holographic phase diagrams. This is currently under investigation. 

\vspace{1cm}

{\bf Acknowledgments} F. Console thanks CAPES for financial support.


\begin{thebibliography}{99}

\bibitem{Maldacena:1997re}
J.~M.~Maldacena: The Large N limit of superconformal field theories and supergravity,
Adv. Theor. Math. Phys. \textbf{2} (1998), 231. 

\bibitem{gauge_gravity} see e.g. O. Aharony, S. S. Gubser, J. M. Maldacena, H. Ooguri and Y. Oz: Large N field theories, string theory and gravity, Phys. Rept. {\bf 323} (2000) 183; E. D’Hoker and D. Z. Freedman: Supersymmetric gauge theories and the AdS/CFT correspondence,
arXiv:hep-th/0201253; M. Benna and I. Klebanov: Gauge-string duality and some applications, arXiv:
0803.1315 [hep-th].


\bibitem{hhh} S.~A.~Hartnoll, C.~P.~Herzog and G.~T.~Horowitz: Building a Holographic Superconductor,
Phys. Rev. Lett. \textbf{101} (2008), 031601; S.~A.~Hartnoll, C.~P.~Herzog and G.~T.~Horowitz: Holographic Superconductors, JHEP \textbf{12} (2008), 015; G.~T.~Horowitz and M.~M.~Roberts, Holographic Superconductors with Various Condensates,
Phys. Rev. D \textbf{78} (2008), 126008.

\bibitem{Horowitz:2010jq}
G.~T.~Horowitz and B.~Way, Complete Phase Diagrams for a Holographic Superconductor/Insulator System,
JHEP \textbf{11} (2010), 011.



\bibitem{Coleman}
S.~R.~Coleman: Q-balls,
Nucl. Phys. B \textbf{262} (1985), 263.

\bibitem{Kusenko}
A.~Kusenko: Small Q balls,
Phys. Lett. B \textbf{404} (1997), 285.

\bibitem{Copeland}
E.~J.~Copeland and M.~I.~Tsumagari: Q-balls in flat potentials,
Phys. Rev. D \textbf{80} (2009), 025016. 

\bibitem{Kaup}
D.~J.~Kaup: Klein-Gordon Geon,
Phys. Rev. \textbf{172} (1968), 1331.

\bibitem{Friedberg}
R.~Friedberg, T.~D.~Lee and Y.~Pang: Scalar Soliton Stars and Black Holes,
Phys. Rev. D \textbf{35} (1987), 3658.


\bibitem{Jetzer}
P.~Jetzer: Boson stars,
Phys. Rept. \textbf{220} (1992), 163.


\bibitem{Schunck}
F.~E.~Schunck and E.~W.~Mielke: General relativistic boson stars,
Class. Quant. Grav. \textbf{20} (2003), R301.


\bibitem{radu}
D.~Astefanesei and E.~Radu: Boson stars with negative cosmological constant,
Nucl. Phys. B \textbf{665} (2003), 594. 


\bibitem{Hartmann_Riedel}
B.~Hartmann and J.~Riedel: Glueball condensates as holographic duals of supersymmetric Q-balls and boson stars,
Phys. Rev. D \textbf{86} (2012), 104008. 


\bibitem{cbs1} P. Jetzer and J. J. van der Bij: Charged boson stars, Phys. Lett. B 227 (1989) 341.

\bibitem{cbs2} H. Arodz and J. Lis: Compact Q-balls and Q-shells in a scalar electrodynamics, Phys. Rev. D 79, 045002 (2009).

\bibitem{cbs3} B. Kleihaus, J. Kunz, C. L\"ammerzahl and M. List: Charged Boson Stars and Black Holes, Phys. Lett. B 675 (2009) 102.


\bibitem{cbs4}
Y.~Brihaye, V.~Diemer and B.~Hartmann: Charged Q-balls and boson stars and dynamics of charged test particles,
Phys. Rev. D \textbf{89} (2014), 084048. 

\bibitem{gauged_BS_power_law}
Y.~Brihaye and B.~Hartmann: Boson stars and black holes with wavy scalar hair,
Phys. Rev. D \textbf{105} (2022) no.10, 104063.

\bibitem{cbs5} 
D.~Pugliese, H.~Quevedo, J.~A.~Rueda H. and R.~Ruffini: On charged boson stars, Phys. Rev. D \textbf{88} (2013), 024053. 




\bibitem{Bizon:2011gg}
P.~Bizon and A.~Rostworowski: On weakly turbulent instability of anti-de Sitter space,
Phys. Rev. Lett. \textbf{107} (2011), 031102.

\bibitem{Dias:2012tq}
O.~J.~C.~Dias, G.~T.~Horowitz, D.~Marolf and J.~E.~Santos: On the Nonlinear Stability of Asymptotically Anti-de Sitter Solutions,
Class. Quant. Grav. \textbf{29} (2012), 235019.

\bibitem{Brihaye:2014bqa}
Y.~Brihaye, B.~Hartmann and J.~Riedel: Self-interacting boson stars with a single Killing vector field in anti\textendash{}de Sitter space-time,
Phys. Rev. D \textbf{92} (2015) no.4, 044049. 





\end{thebibliography}
\end{document}